\documentclass[a4paper,fleqn,useAMS,usenatbib]{mnras}

\pdfoutput=1

\usepackage{natbib}
\usepackage{mathptmx}

\usepackage[T1]{fontenc}
\usepackage{ae,aecompl}

\usepackage[dvips]{graphicx}
\usepackage{amsmath}
\usepackage{amsfonts}
\usepackage{amssymb}
\usepackage{comment}
\usepackage{bm} 
\usepackage[dvipsnames]{xcolor}
\usepackage[]{hyperref}
\hypersetup{
	colorlinks=true,	
	urlcolor=MidnightBlue,		
	hypertexnames=true,
	plainpages=false,
	naturalnames=true,
	pdftitle={Binary deviations from single object astrometry},    
	pdfauthor={Zephyr Penoyre},     
	linkcolor=WildStrawberry,          
	citecolor=Periwinkle,        
}

\usepackage{amsmath}
\usepackage{amsfonts}
\usepackage{amssymb}
\usepackage{wasysym}
\usepackage[percent]{overpic}
\usepackage{subfig}

\makeatletter
\def\env@matrix{\hskip -\arraycolsep 
  \let\@ifnextchar\new@ifnextchar
  \array{*{\c@MaxMatrixCols}c}}
\makeatother

\title[Binary deviations]{Binary deviations from single object astrometry}
\author[Z. Penoyre et al.]{Zephyr Penoyre$^{1}$\thanks{E-mail:
\href{mailto:zpenoyre@ast.cam.ac.uk}{zpenoyre@ast.cam.ac.uk}}, Vasily Belokurov$^{1}$, N. Wyn Evans$^{1}$, A. Everall$^{1}$, S.~E. Koposov$^{2,1,3}$ \\
$^{1}$Institute of Astronomy, University of Cambridge, Madingley Road, Cambridge, CB3 0HA, United Kingdom \\
$^{2}$McWilliams Center for Cosmology, Carnegie Mellon University, 5000 Forbes Ave, Pittsburgh, PA 15213, USA\\
$^{3}$Kavli
 Institute for Cosmology, University of Cambridge, Madingley Road, Cambridge CB3 0HA, UK}
\date{Accepted . Received ; in original form }

\pubyear{2019}

\begin{document}
\label{firstpage}
\pagerange{\pageref{firstpage}--\pageref{lastpage}}
\maketitle

\begin{abstract}
Most binaries are undetected. Astrometric reductions of a system using the assumption that the object moves like a single point mass can be biased by unresolved binary stars. The discrepancy between the \textit{centre of mass} of the system (which moves like a point mass) and the \textit{centre of light} (which is what we observe) introduces additional motion. We explore the extent to which binary systems affect single object models fit to astrometric data. This tells us how observations are diluted by binaries and which systems cause the largest discrepancies - but also allows us to make inferences about the binarity of populations based on observed astrometric error. By examining a sample of mock observations, we show that binaries with periods close to one year can mimic parallax and thus bias distance measurements, whilst long period binaries can introduce significant apparent proper motion. Whilst these changes can soak up some of the error introduced by the binary, the total deviation from the best fitting model can be translated into a lower limit on the on-sky separation of the pair. Throughout we link these predictions to data from the {\it Gaia} satellite, whilst leaving the conclusions generalizable to other surveys.
\end{abstract}

\begin{keywords}
astrometry,
parallaxes,
proper motions,
binaries: general,
methods: analytical
\end{keywords}

\section{Introduction}

Around half of the solar type stars are in binary systems \citep{DM1991,Raghavan10}. Amongst more massive stars, the fraction is higher still \citep{Sana2012,Duchene13}. The number and properties of binaries tell us about the conditions needed for star formation \citep{Shu1987, Bate1995,Bonnell1998}, the properties of the disks in which they form, and their motions both in the dense stellar nursery and during their long life-histories. Binarity can affect the formation of planets and the eventual fate of a star. Some of the most exotic objects in the Galaxy result from the evolution of binary systems -- including cataclysmic variables, hot-Jupiters and type 1a supernova \citep{Whelan1973,Tutukov1981,Webbink1984}. It is even possible for a bright binary star companion to tell us of the presence of a massive dark body, such as a cool white dwarf or a stellar mass black hole \citep{Andrews19}.

However, the number of known binaries is tiny compared to the total number of catalogued stars. By many metrics, a typical binary star appears indistinguishable from a single object. This means that most binaries on-sky are still undetected, and the information that can be gleaned from them still ready to be reaped. But it also means that, though they may not be detectable, any observation we make assuming all stars are singular will necessarily be contaminated by binaries.

Currently, we know of around a hundred thousand visual binaries \citep{Hartman20}, in which both objects can be separately resolved. Eclipsing binaries, in which the system is by chance aligned such that one star partially occults the other, are numbered in the thousands \citep{Prsa11,Kirk16}.
Spectroscopic binaries, whose motion can be seen by measuring the shift in the radial velocities over an orbit, are growing in number with many thousands now known \citep{Price-Whelan20}. These probe close binary pairs (generally with periods of a year or less), for which multiple periods can be observed and which cause measurably large radial velocity shifts.

Instruments like {\it Hipparcos} \citep{Perryman97} and {\it Gaia} \citep{Gaia16} measure the position of stars on sky to very high accuracy and enable the discovery of astrometric binaries. These show significant binary motion on sky, on top of the motion of the centre of mass, and are becoming detectable as the noise floor of astrometric observations reaches milliarcseconds ($mas$) precision \citep{Gaia18}. Long period systems will cause an extra component of motion of the centre of mass, changing slightly over time, termed a proper motion anomaly \citep{Kervella19}, which is sensitive to relatively long period binaries (around 10 years or more).

In this paper, we focus on another way in which binarity can impinge on astrometric measurements: as excess error on a single-body astrometric fit. For systems with periods less than $\sim 10$ years, a significant ($\gtrsim 1/2$) number of orbits can be completed over the observing period of a single instrument. Thus, we expect the astrometric motion of the centre of mass to be well captured, but the binary motion to cause excess noise, which could potentially be harvested as inferred properties of the stars. By examining the analytic deviations in an ideal case (Section \ref{sec:analytic}), we can make predictions which we then compare to a fuller numerical model (Section \ref{sec:numerical}) and mock observations (Section \ref{sec:mock}). We can then gain insight into when the analytic treatment is accurate, and for which types of systems we expect significant deviation. Finally, in Section \ref{sec:observe}, we discuss the impact of these effects on observable properties, both as an identifier of binary systems and a nuisance signal.

This analysis is heavily motivated by the {\it Gaia} survey, a space-based telescope measuring the position and velocities of millions of stars at a precision of a few milliarcseconds ($mas$) over a period of years. However, most of the behaviour is general, and could be applied to surveys before and after {\it Gaia}. Thus, we will attempt to remain general, and invoke the properties of the {\it Gaia} survey only when fiducial numerical values are needed to further explore the results.

\section{Analytic deviations}
\label{sec:analytic}

In the limit that the period of a binary is $\lesssim$ the observing time of a survey, we can derive an accurate analytic description of the magnitude of on-sky deviation between the movement of the centre of light compared to the centre of mass.

\subsection{Offset between the centre of mass and of light}
\label{sec:offset}

Let us start with a binary system, in which we label the bodies A and B. We assume that A is the more luminous of the two. Thus, we can remap the brightnesses of the two objects onto an absolute luminosity and a ratio such that $L=L_A$ and $l={L_B}/{L_A}$. The total luminosity is $L_A+L_B = L(1+l)$ with $l<1$. We can define the mass similarly, with $M=M_A$ and $q={M_B}/{M_A}$, which can be less than or greater than one. For a value significantly greater than one, the system can be described in the simpler framework of a massive dark companion \citep{Shahaf19}.

When we do not spatially resolve a binary, or -- to be more precise -- we cannot separate the point spread function of the two sources, we see the system at the position of its centre of light (c.o.l). However, the dynamics of the system govern the motion of the centre of mass (c.o.m.) and thus if these two are offset, the c.o.l. will orbit around the c.o.m. causing the system to appear to be moving non-inertially.

If the total distance between the two sources is $d$, then the distance from A to the c.o.m. is $d q/(1+q)$ while the distance to the c.o.l. is $d l/(1+l)$\footnote{This is derived and explored in more detail in appendix \ref{ap:col}}. Thus, if $\delta d$ is the physical distance between the two, the fraction of the distance between the c.o.m. and the c.o.l. is 
\begin{equation}
\Delta = \frac{\delta d}{d}= \frac{|q-l|}{(1+q)(1+l)}
\end{equation}

\subsection{Two-body orbits}

Assuming Keplerian potentials and no external forces, the orbit follows the usual parametric form. Using the eccentric anomaly $\eta$, the evolution of the separation $d$ with time $t$ is
\begin{equation}
\label{orbitEq}
d = a(1-e\cos{\eta}),\qquad\qquad
t=\frac{P}{2\pi}(\eta - e \sin{\eta}).
\end{equation}
The orbital phase $\phi$ satisfies
\begin{equation}
\label{eq:trigfuncs}
\cos{\phi} = \frac{\cos{\eta} - e}{1-e\cos{\eta}}
\qquad\qquad
\sin{\phi} = \frac{\sqrt{1-e^2} \sin{\eta}}{1-e\cos{\eta}}.
\end{equation}
The semi-major axis $a$ and period $P$ are
\begin{equation}
a= -\frac{q GM^2}{2E},\qquad\qquad
\frac{P}{2\pi} = \sqrt{\frac{a^3}{(1+q)GM}}.
\end{equation}
and thus specifying $M$ and $q$ and either of $a$ or $T$ is sufficient to calculate the other.
The eccentricity is
\begin{equation}
e^2=1- \frac{1+q}{q^2}\frac{L^2}{GM^3 a}.
\end{equation}
Here $E (<0)$ and $L$ are the total energy and angular momentum of the two bodies and are constant over the orbit. We cannot solve eq~(\ref{orbitEq}) directly for $r(t)$, though approximate solutions are possible \citep{Penoyre19}.


\subsection{On-sky projection}

It will be useful to work in spherical coordinates $(r,\theta,\phi)$ where $\theta$ is the polar angle ranging from $0$ to $\pi$ and $\phi$ is the azimuthal angle ranging from $0$ to $2\pi$. We are free to set the orientation and thus align it with the phase of the binary at periapse (thus making $\phi$ both the orbital phase and the azimuthal coordinate) and the origin is at the c.o.m. of the binary.

The two components of the binary lie in the $\pm \mathbf{\hat{d}} $ directions where
\begin{equation}
\label{dHat}
\mathbf{\hat{d}} =\begin{pmatrix} \cos{\phi} \\ \sin{\phi} \\ 0\end{pmatrix}
\end{equation}
in Cartesian coordinates.

We can also define the position of an observer in this frame by two angles, $\theta_v$ (equivalent to inclination, $i$) and $\phi_v$ such that the vector pointing towards the observer along the line of sight is
\begin{equation}
\mathbf{\hat{l}}=\begin{pmatrix} \cos{\phi_v}\sin{\theta_v} \\ \sin{\phi_v}\sin{\theta_v} \\ \cos{\theta_v}\end{pmatrix}.
\end{equation}

Figure \ref{sketch} gives a quick sketch of the system and the coordinates used, which we must transform to an on-sky projection and eventually to a motion in a specific coordinate frame (e.g. ecliptic longitude and latitude or RA and Dec).

\subsection{Deviations for a circular binary orbit}

In the case of a circular orbit, the formulae thus presented are sufficient to describe the deviations caused by the binary. The magnitude of the projected binary separation is
\begin{equation}
|s| = d \left|\mathbf{\hat{l}} \wedge \mathbf{\hat{d}} \right|= d\sqrt{1-\sin^2\theta_v ( 1- \sin^2\varphi_v)},
\end{equation}
where $\varphi_v = \phi - \phi_v$ is the azimuthal angle between the orbital phase and the line-of-sight vector.

Let $\boldsymbol{\epsilon}$ be the 2D deviations of the c.o.l. from c.o.m. in on-sky coordinates - where we will express this here as an angle by multiplying through by the parallax of the object $\varpi$ (this assumes $a$ is expressed in $AU$ and thus $\epsilon$ has the same units as $\varpi$). The physical distance, projected perpendicular to the line-of-sight, between the c.o.m. and c.o.l. is thus
\begin{equation}
\label{eq:epsilon}
|\boldsymbol{\epsilon}| = \Delta \varpi |s| = \frac{\varpi a|q-l|}{(1+q)(1+l)} \frac{1-e^2}{1+e\cos\phi}\sqrt{1-\sin^2\theta_v ( 1- \sin^2\varphi_v)}.
\end{equation}
Assuming that many periods of the binary are observed, the inferred astrometric scatter of a single-body fit is equal to \begin{equation}
\label{dThetaGen}
\delta\theta=\sqrt{\langle|\boldsymbol{\epsilon}-\langle\boldsymbol{\epsilon}\rangle|^2\rangle} = \sqrt{\langle|\boldsymbol{\epsilon}|^2\rangle - |\langle\boldsymbol{\epsilon}\rangle|^2}
\end{equation}
where $\langle \rangle$ denotes the time-average. If we assume a sufficient number of orbits occur over the observing period, we can take this average to be over one-complete orbit. For the circular case $\langle\boldsymbol{\epsilon}\rangle$ is zero, i.e. the average position of the object is at the focii, and thus
\begin{equation}
\label{eq:circEps}
\delta\theta_{circ}=\sqrt{\langle|\boldsymbol{\epsilon}|^2\rangle}=\frac{\varpi a|q-l|}{(1+q)(1+l)}\sqrt{1-\frac{\sin^2\theta_v}{2}}.
\end{equation}
For non-circular orbits $\langle\boldsymbol{\epsilon}\rangle$ is not in general zero and thus we must express the position on-sky fully to calculate $\delta$.

\subsection{Binary motion across the sky-plane}
\label{sec:skyplane}

\begin{figure}
\centering
\includegraphics[width=0.49\textwidth]{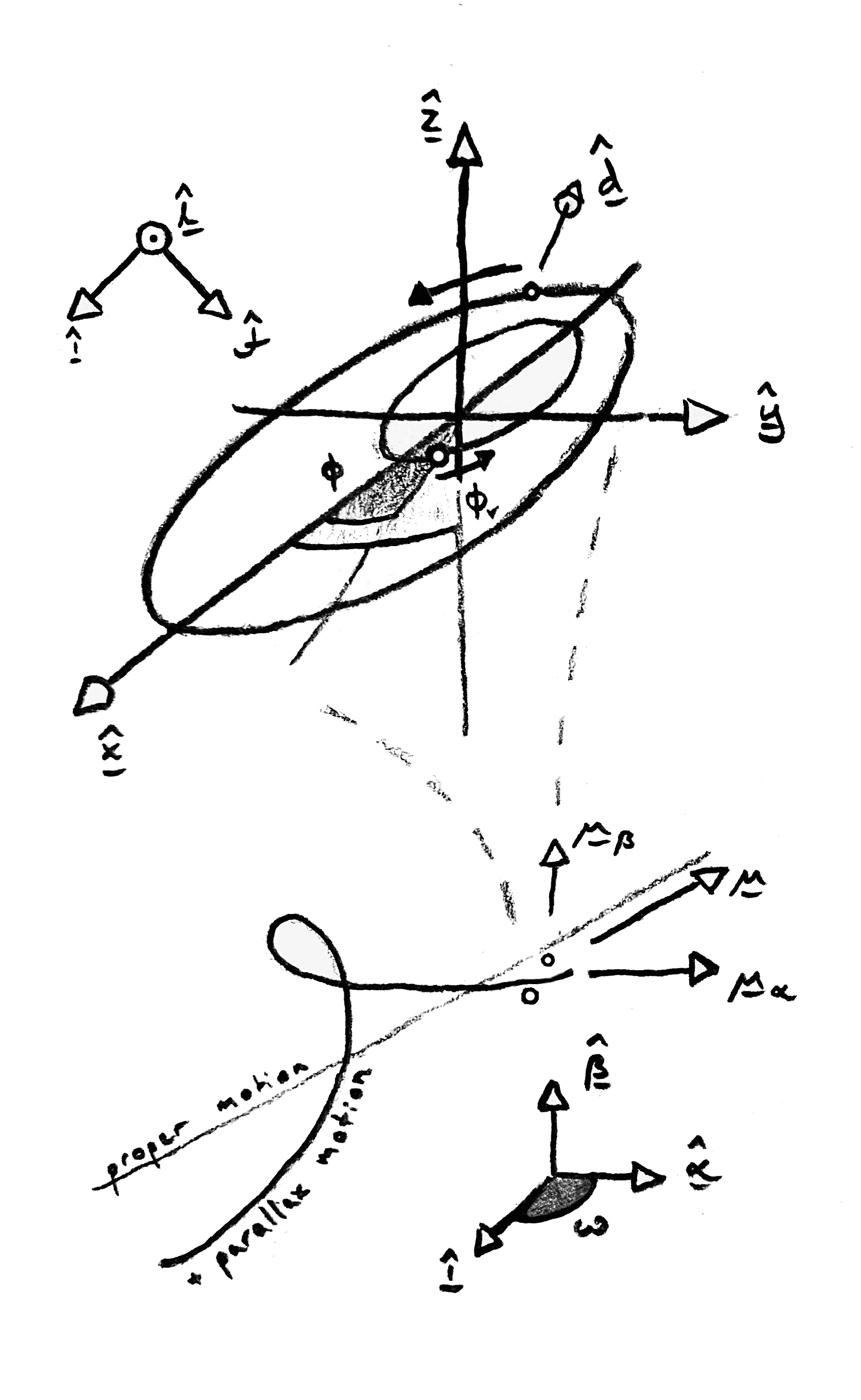}
\caption{Sketch of the coordinate systems used to describe the motion of the binary and the centre of mass. Both views are shown as they would appear on-sky (if the system could be resolved). Upper panel: Binary motion around the centre of mass (at the origin) as described in $(x,y,z)$ coordinates with the orbit confined to the $(x,y)$ plane and periapse of the brighter source on the $x$ axis. The azimuthal viewing angle $\phi_v$ is shown, whilst the polar viewing angle $\theta_v$ cannot be, as it is the angle between the vectors $\mathbf{\hat{z}}$ and $\mathbf{\hat{l}}$. The orbital phase of the binary is given by $\phi$, and the phase relative to the viewing angle is $\varphi_v=\phi-\phi_v$ (if shown in this diagram it would be almost $2\pi$). The centre of light will sit on the line between the two stars, at a constant fraction of the distance. To convert into on-sky motion, we project along the $\mathbf{\hat{i}},\mathbf{\hat{j}},\mathbf{\hat{l}}$ directions which lie in, and perpendicular to, the plane of the page. Bottom panel: The motion of the centre of mass of the system follows a straight line proper motion across the sky, with added motion caused by parallax (the change of a nearby object's position compared to the background due to Earth's orbit around the Sun). The final coordinate conversion is from on-sky coordinates with arbitrary direction specified by $\omega$ to those which line up with our reference coordinate system (e.g. ecliptic longitude and latitude or RA and Dec).}
\label{sketch}
\end{figure}

It is natural to define two other unit vectors which capture the projection on the plane perpendicular to $\mathbf{\hat{l}}$ (and thus describe the position on-sky). These vectors can be taken as
\begin{equation}
\mathbf{\hat{i}} = \frac{1}{\sqrt{1-\sin^2\theta_v \cos^2\phi_v}}\begin{pmatrix}1-\cos^2\phi_v \sin^2\theta_v \\ -\sin\phi_v \cos \phi_v \sin^2\theta_v \\ -\cos\phi_v \sin \theta_v \cos\theta_v   \end{pmatrix}
\end{equation}
which projects onto the $x$-axis and thus appears to pass through the orbital periapse and apoapse and
\begin{equation}
\mathbf{\hat{j}} = \frac{1}{\sqrt{1-\sin^2\theta_v \cos^2\phi_v}}\begin{pmatrix} 0 \\  \cos \theta_v \\ -\sin\phi_v \sin \theta_v \end{pmatrix}
\end{equation}
which completes the orthonormal set. The choice of these two directions is arbitrary (at least until we define the orientation of the system on the sky), but provides an easily interpretable coordinate system.

In these coordinates, the centre of light at time $t$ is at position
\begin{equation}
\begin{aligned}
i=&\Delta(\mathbf{\hat{i}} \cdot \mathbf{d})= \frac{\Delta a(1-e^2)}{1+e\cos\phi}\frac{\cos\phi-\cos\varphi_v \cos\phi_v \sin^2\theta_v}{\sqrt{1-\sin^2\theta_v \cos^2\phi_v}}, \\
j=&\Delta(\mathbf{\hat{j}} \cdot \mathbf{d})= \frac{\Delta a(1-e^2)}{1+e\cos\phi}\frac{\sin\phi \cos\theta_v}{\sqrt{1-\sin^2\theta_v \cos^2\phi_v}}, \\
l=&\Delta(\mathbf{\hat{l}} \cdot \mathbf{d})= \frac{\Delta a(1-e^2)}{1+e\cos\phi}\cos\varphi_v \sin\theta_v
\end{aligned}
\end{equation}
where $\varphi_v=\phi-\phi_v$ is the azimuthal angle between the current position and the line of sight and $\mathbf{d}$ is specified by eqs (\ref{orbitEq}) and (\ref{dHat}).

Thus, at any given time, the full on-sky deviation from the c.o.l. to c.o.m. is given by
\begin{equation}
\label{epsVectorShort}
\boldsymbol{\epsilon}=\varpi \begin{pmatrix} i \\  j \end{pmatrix}.
\end{equation}
Taking the time average of this, we find
\begin{equation}
\langle \boldsymbol{\epsilon} \rangle = -\frac{3e\varpi |q-l| a}{2(1+q)(1+l)}\sqrt{1-\cos^2\phi_v\sin^2\theta_v}\begin{pmatrix} 1 \\  0 \end{pmatrix}
\end{equation}
and thus the full expression for $\delta \theta$ given any eccentricity is
\begin{equation}
\label{deltaFull}
\delta\theta = \frac{\varpi a|q-l|}{(1+q)(1+l)}\sqrt{1-\frac{\sin^2\theta_v}{2}-\frac{3+\sin^2\theta_v(\cos^2\phi_v-2) }{4} e^2 }.
\end{equation}
In the circular limit, this agrees with the expression in eq~(\ref{eq:circEps}) as expected. For orbits approaching radial, we find
\begin{equation}
\delta\theta_{rad} = \frac{\varpi a|q-l|}{2(1+q)(1+l)}\sqrt{1-\cos^2\phi_v \sin^2\theta_v}.
\end{equation}
Inspection of equation \ref{deltaFull} shows that eccentricity always decreases $\delta \theta$.

In Appendix \ref{ap:anLong}, we derive a comprehensive, but significantly more complex, expression for the full behaviour.

\subsection{Binary motion as astrometric error}

If we do not resolve a binary and fit it as a single source, this on-sky motion of the centre of light behaves as a source of astrometric error in two ways.

First, it may change the inferred astrometric parameters. Significant motion due to a binary could be interpreted as a change in the motion of a single point source. Long period binaries will have a roughly constant offset from the centre of mass. Intermediate-period systems, which complete a significant fraction of one orbit over the observing time, can mimic significant proper motion. We investigate this in detail for our numerical models in Section~\ref{sec:numerical}. Secondly, the motion of systems with periods close to a year may mimic parallactic motion, and thus bias our estimate of the distance. Thirdly, it may increase the observed noise. For systems for which the astrometric fit to the c.o.l. motion well matches the c.o.m. motion, any on-sky deviations appear as additional noise. Thus, systems with anomalously high inferred noise may be caused by binaries -- and conversely high observed noise can be translated to a prediction of binary properties. 

Apparent motion of a single star on the sky, though it has five free-parameters, is highly constrained -- consisting of straight line motion across the sky superimposed on the apparent motion of nearby sources due to Earth's orbit around the Sun. The form of the latter is fixed by the source position on-sky, with the parallax only scaling the magnitude of the effect. Thus, it is hard, though far from impossible (especially in huge datasets such as the {\it Gaia} survey), to convincingly mimic astrometric motion with short period binaries. Thus, binary motion will mostly contribute to increased error in these cases.

Any observation also has some intrinsic astrometric error, $\sigma_{\rm ast}$ (which is in part a function of the source, but we shall take as constant for a given instrument). 

When we fit an astrometric model onto a single source, we can construct the statistic
\begin{equation}
\chi^2 = \sum^{N_{\rm obs}} \left(\frac{(\alpha_{\rm obs}-\alpha_{\rm model})^2 + (\beta_{\rm obs}-\beta_{\rm model})^2}{\sigma_{\rm ast}^2} \right)
\end{equation}
which will be $\sim N_{\rm obs}-5$ for a single well-behaved source. Here $\alpha$ and $\beta$ are the on-sky angular coordinates and $N_{\rm obs}$ is the number of observations of the source, and the five corresponds to the degrees of freedom of the astrometric fit.

When we add the induced motion of a binary, we expect the $\chi^2$ to increase by $\chi^2_{\rm binary}=\delta \theta^2/ \sigma_{\rm ast}^2$ from which we can make a prediction of the Unit Weight Error (UWE)
\begin{equation}
\label{uwedTheta}
{\rm UWE}_{\rm pred} = \sqrt{\frac{\chi^2_{\rm total}}{N_{\rm obs}-5}} \simeq \sqrt{1+\left(\frac{\delta \theta}{\sigma_{\rm ast}}\right)^2}
\end{equation}
(ignoring cross-terms).


This is a parameter recorded by the {\it Gaia} survey, in the form of Renormalised Unit Weight Error (RUWE) rescaling to account for systematic trends in the measurements\footnote{See the {\it Gaia} consortium's technical note \href{http://www.rssd.esa.int/doc_fetch.php?id=3757412}{GAIA-C3-TN-LU-LL-124-01}} and thus can be used to explore the distribution of binaries across the huge {\it Gaia} DR2 catalog (Belokurov et al. 2020, submitted).

\section{Numerical deviations}
\label{sec:numerical}

For systems with longer periods, or those inhabiting certain (un)fortunate parts of parameter space, the presence of a binary companion biases the astrometric fit, such that the inferred parallax, position and proper motion are inaccurate. These cases defy easy analytical exploration, but can be modelled numerically. By fitting single body astrometric solutions to their more complex on-sky motion, we can compare the accuracy and precision of our astrometric fits as a function of the properties of the binary.

\subsection{Path of the centre of mass}

Given the above tools, we map out the path across the sky of the centre of light of a binary system - combining the on-sky motion of the centre of mass and the centre of light. The former moves as a single body, including the effects of parallax, and can travel a non-negligible distance on-sky over the observing period. The latter is a correction to this motion capturing the Keplerian binary orbit, and can be assumed to be sufficiently small such that the orbit is constant over the observing period (i.e. the parameters of the orbit, including viewing angles, are constant) and a linear correction to the centre of mass position.

The single body motion depends on the orbit of the Earth, and the position and proper motion. The most natural coordinate system to use is the ecliptic, as it is the Earth's motion around the Sun that traces the parallactic ellipse. Letting $\phi_E$ be the phase of earth's orbit and  $e_E$ the eccentricity, we can express the full single body astrometric motion as
\begin{equation}
\begin{aligned}
\label{mdalpha}
\Delta\alpha(t)=&\Delta\alpha_0+\left(t-t_0-t_b\right)\mu_\alpha \\ &- \frac{\varpi}{\cos\beta_i}\left(\cos\psi + e_E(\sin\psi\sin\tau-\cos\phi)\right)
\end{aligned}
\end{equation}
and
\begin{equation}
\begin{aligned}
\label{mdbeta}
\Delta\beta(t)=&\Delta\beta_0+\left(t-t_0-t_b\right)\mu_\beta \\ &- \varpi\sin\beta_i\left(\sin\psi + e_E(\cos\psi\sin\tau+\sin\phi)\right)
\end{aligned}
\end{equation}
where
\begin{equation}
t_b=\frac{A_u\cos\beta_i}{c}\left(\cos\psi-\cos\psi_0 + e_E(\sin\tau\sin\psi-\sin\tau_0\cos\psi_0)\right),
\end{equation}
$\psi(t)=\phi_E(t)-\tau$ and $\tau=\frac{2\pi(t-t_p)}{T_E}$ where $T_E$ is one year and $t_p$ the time of any pericentre passage of Earth\footnote{For example in relevance to {\it Gaia} we might use $t_p=2456662.00 \ BJD$, shortly before the beginning of astrometric observations in {\it Gaia} DR2, $t_0=2456863.94 \ BJD$}. These results are derived in detail in Appendix \ref{ap:astrometric}.

\subsection{Path of the centre of light}

As the binary separations are always small (compared to angles spanning the whole sky) the addition of the binary component is approximately linear.

Section \ref{sec:skyplane} maps out the contribution due to the binary orbit but one final transformation must be made to align the orientation of the binary system with our on-sky coordinates of choice. We could imagine taking the system shown in Fig.~\ref{sketch} and putting a pin through the origin, along the line of sight, and then rotating the page beneath that pin. This introduces one last viewing angle, the orientation of the system relative to our reference axes, $\omega_v$. We can also at this point move from coordinates describing physical distances to movements on-sky by multiplying through by the parallax, $\varpi$, (for $a$ is given in AU) to give the deviations of the c.o.l. in the azimuthal and polar coordinates of our chosen astronomical system $(\alpha_b,\beta_b)$:
\begin{equation}
\label{raDecBinary}
\alpha_b = \varpi(i \cos\omega_v + j \sin\omega_v) \ \ \mathrm{and} \ \ \beta_b = \varpi(\cos\omega_v j - \sin\omega_v i).
\end{equation}

Thus, adding eqns (\ref{mdalpha}), (\ref{mdbeta}) and (\ref{raDecBinary}), we can describe the motion of the centre of light as observed by a survey such as {\it Gaia}.

Examples of such motions are shown for eight mock observations in Fig.~\ref{binaries}, as detailed in the next section.

\section{Mock observations}
\label{sec:mock}

\begin{figure*}
\captionsetup[subfigure]{labelformat=empty}
\centering
    \subfloat[]{\includegraphics[width=0.75\linewidth]{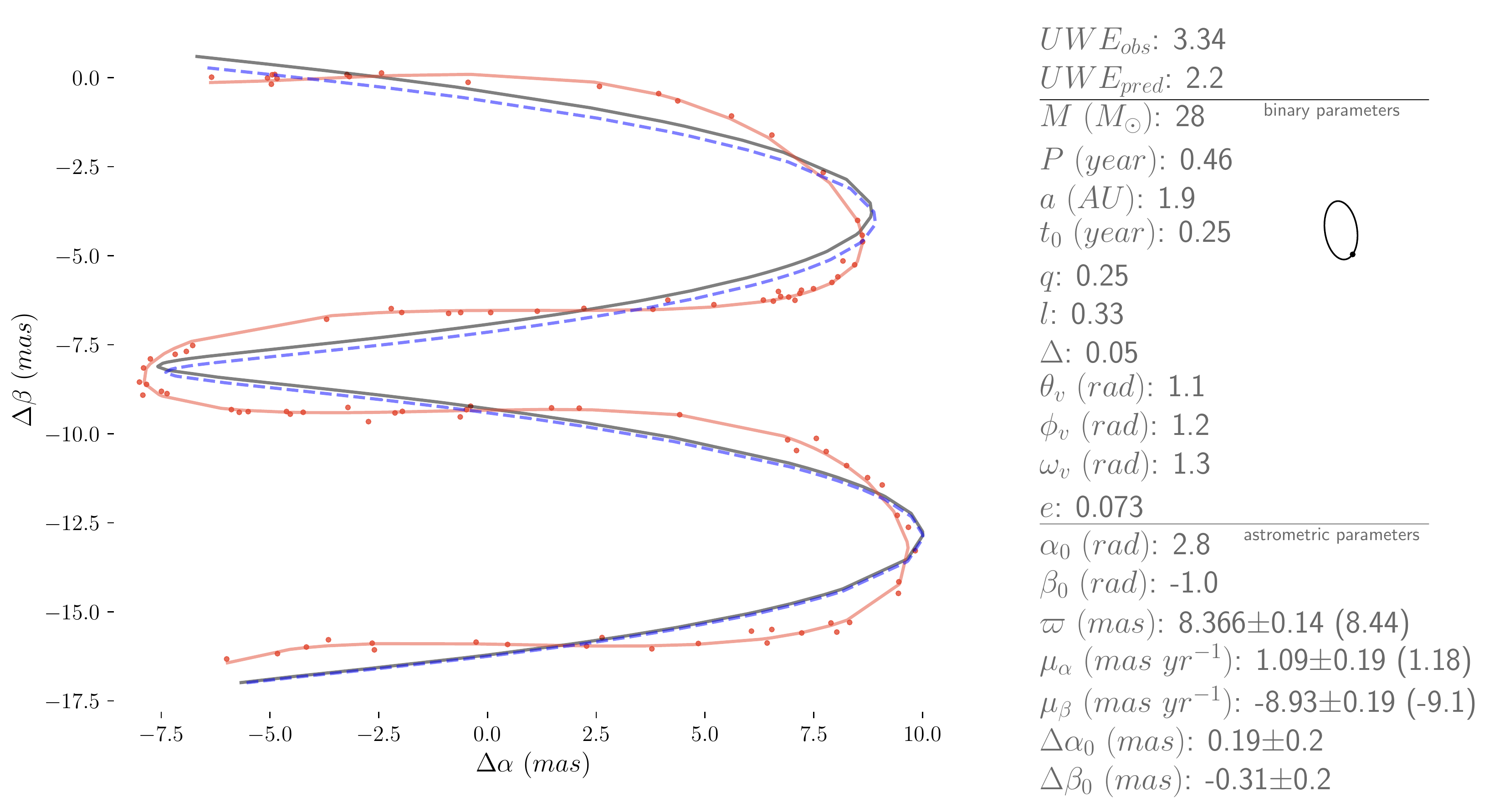}}

    \subfloat[]{\includegraphics[width=0.75\linewidth]{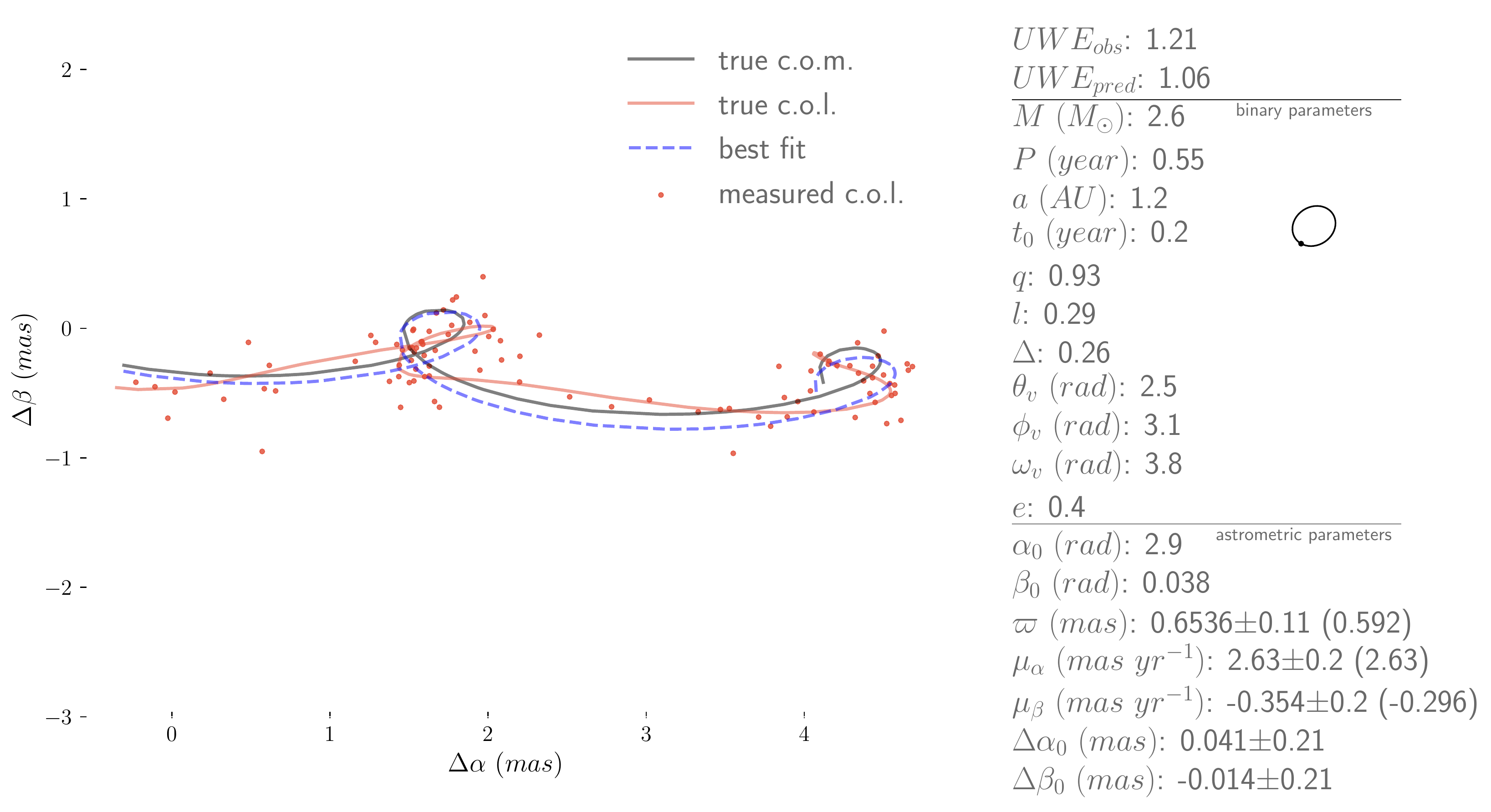}}
\caption{Eight example unresolved binaries, showing: the motion of the centre of mass (black line) which moves as a single body, the motion of the centre of light (red line) which deviates due to binarity, and the simulated observations including error which we fit to (red dots). The best fitting single body curve is shown also (dashed blue). The properties of each system are shown to the right of the plots. Values derived from the least squares fit are given with errors (true values in brackets). Also shown in this inset is the ellipse (or fraction thereof) traced by the centre of light excluding parallax motion - to the same scale as each main panel. One thousand such fits can be viewed  \href{https://drive.google.com/open?id=1uYuBMxkgjr_UVI7SEjHOEWT474OEhAFu}{here}}
\label{binaries}
\end{figure*}

\begin{figure*}
\captionsetup[subfigure]{labelformat=empty}
\ContinuedFloat
\centering

    \subfloat[]{\includegraphics[width=0.7\linewidth]{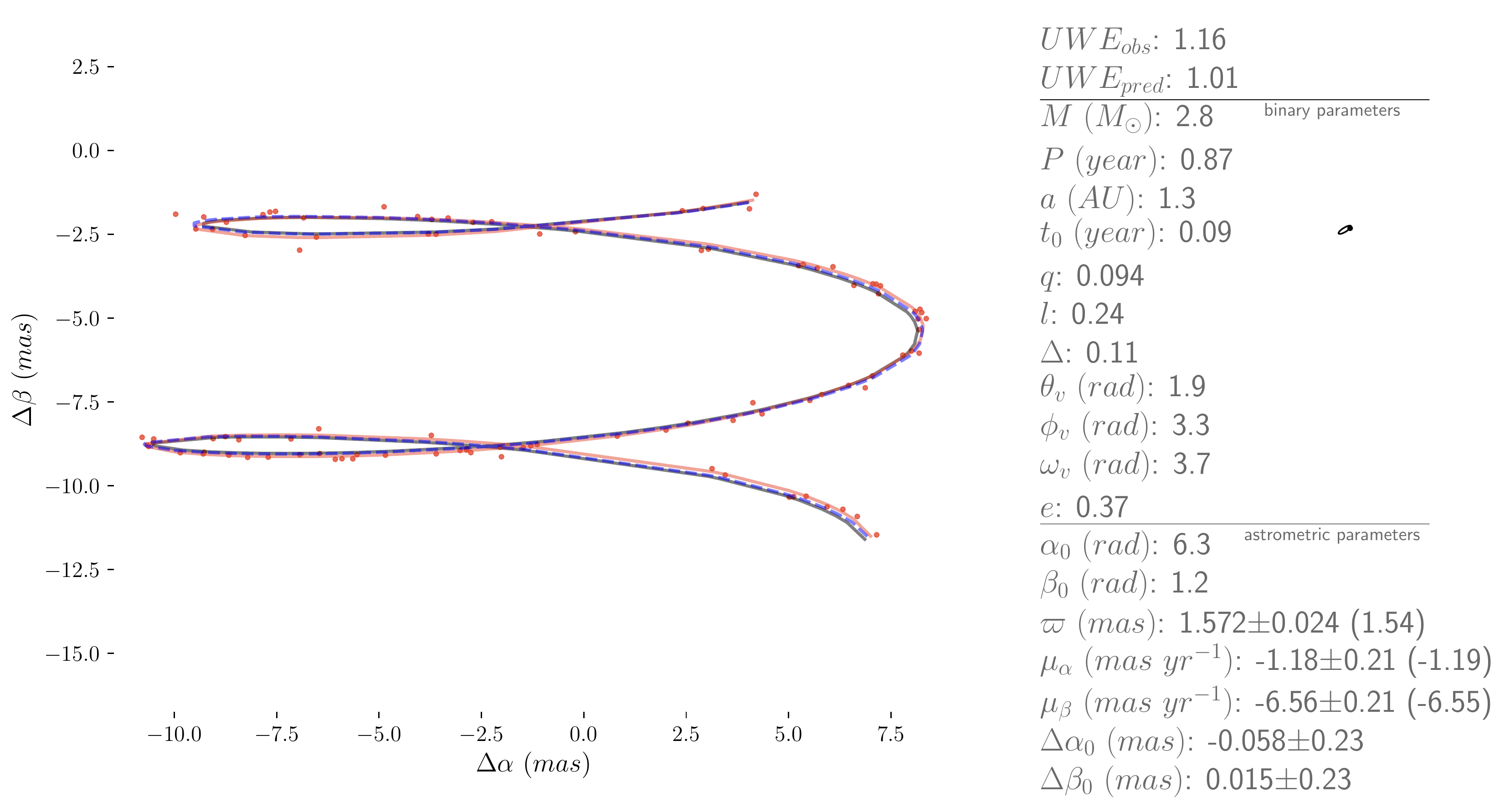}}
    
    \subfloat[]{\includegraphics[width=0.7\linewidth]{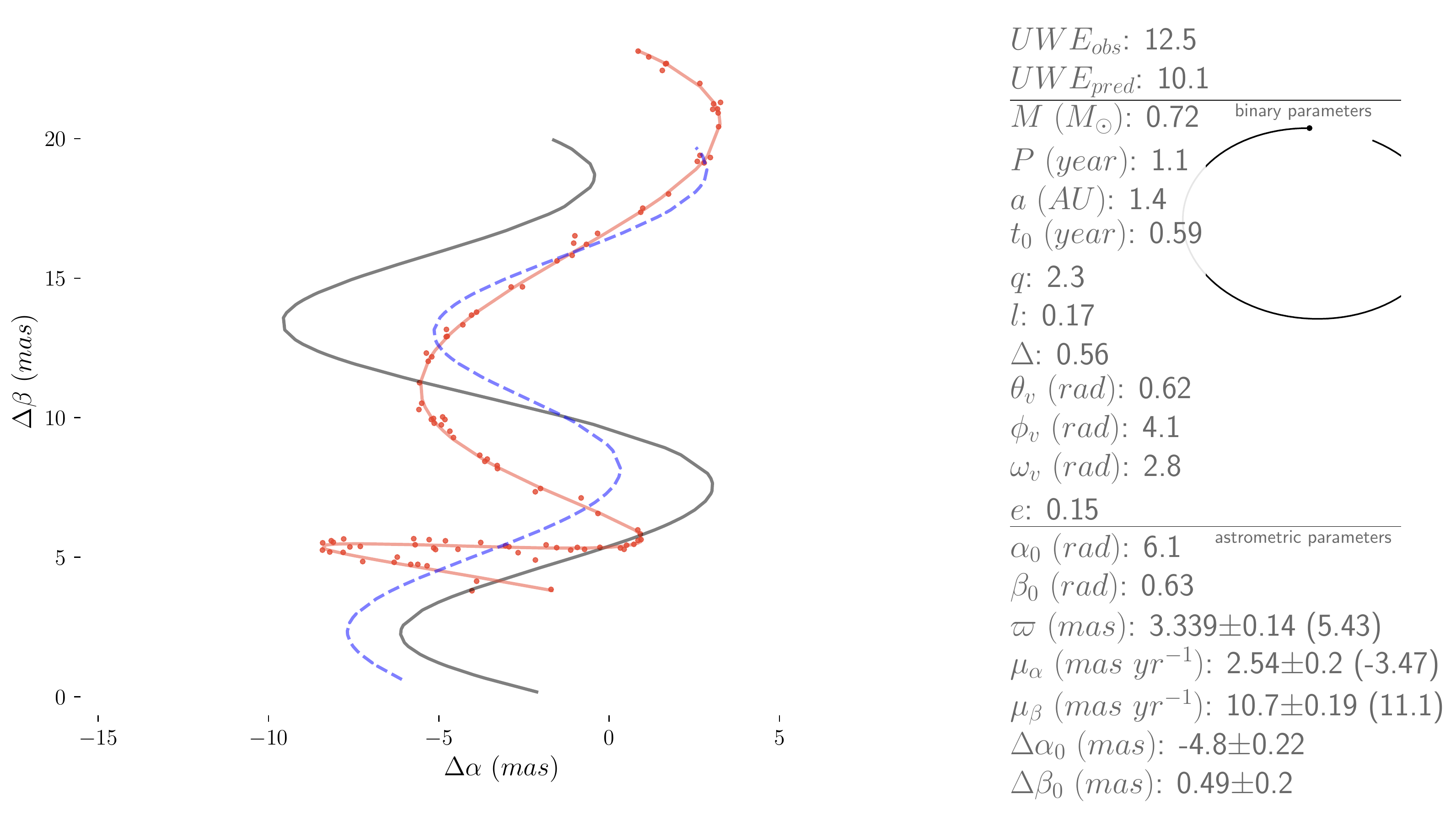}}

    \subfloat[]{\includegraphics[width=0.7\linewidth]{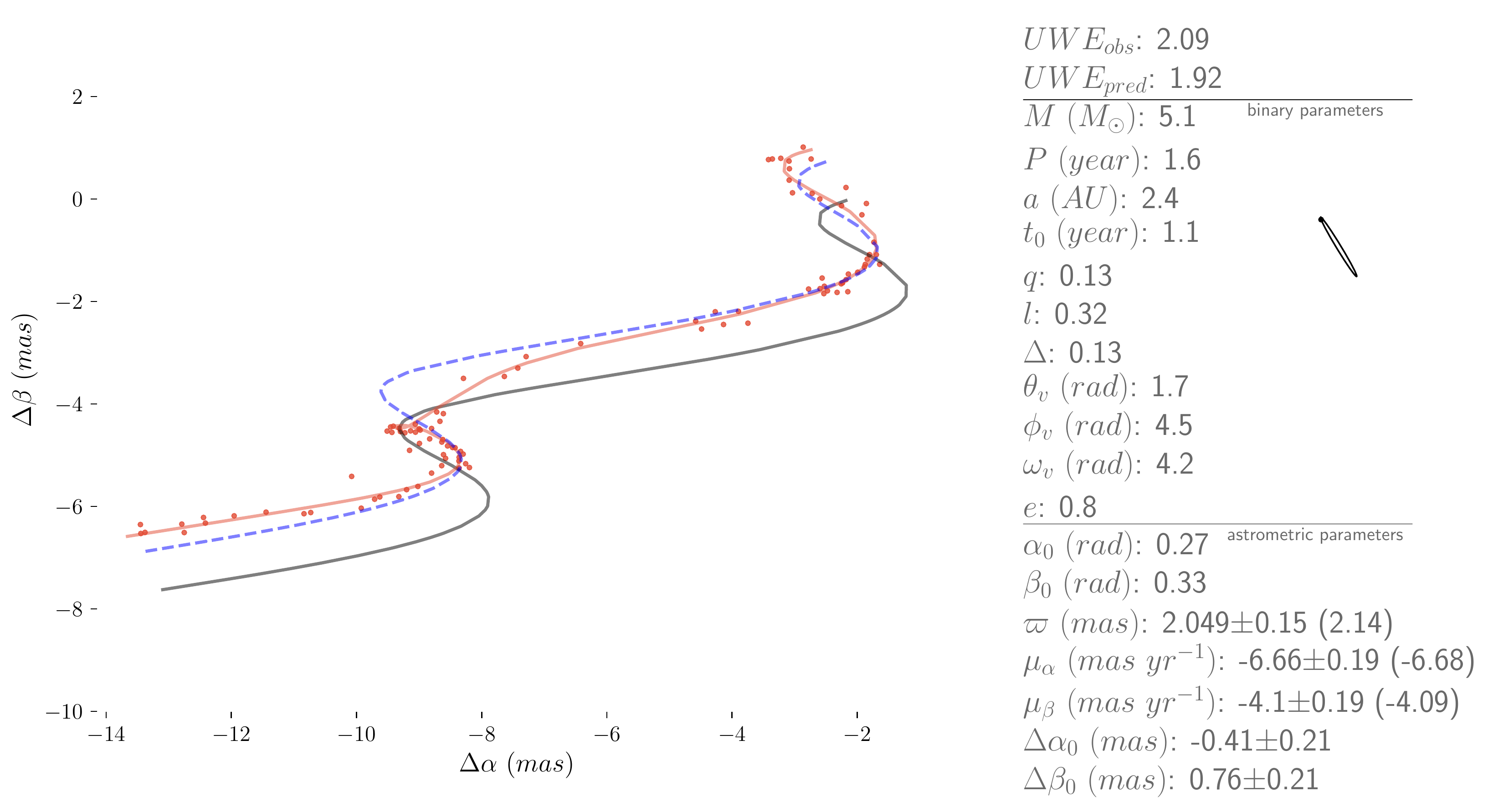}}
\caption[]{(continued)}
\end{figure*}

\begin{figure*}
\captionsetup[subfigure]{labelformat=empty}
\ContinuedFloat
\centering

    \subfloat[]{\includegraphics[width=0.7\linewidth]{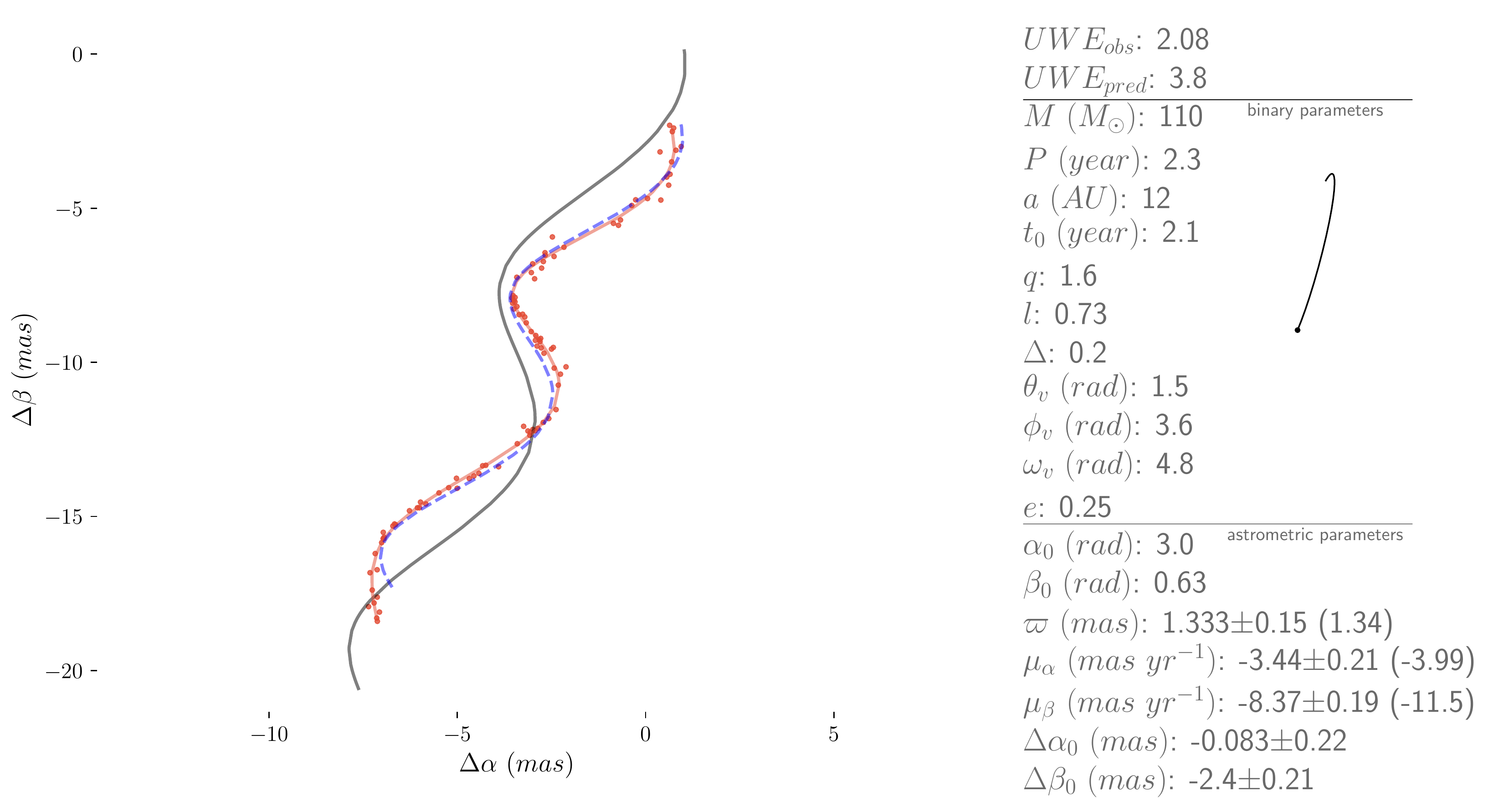}}
    
    \subfloat[]{\includegraphics[width=0.7\linewidth]{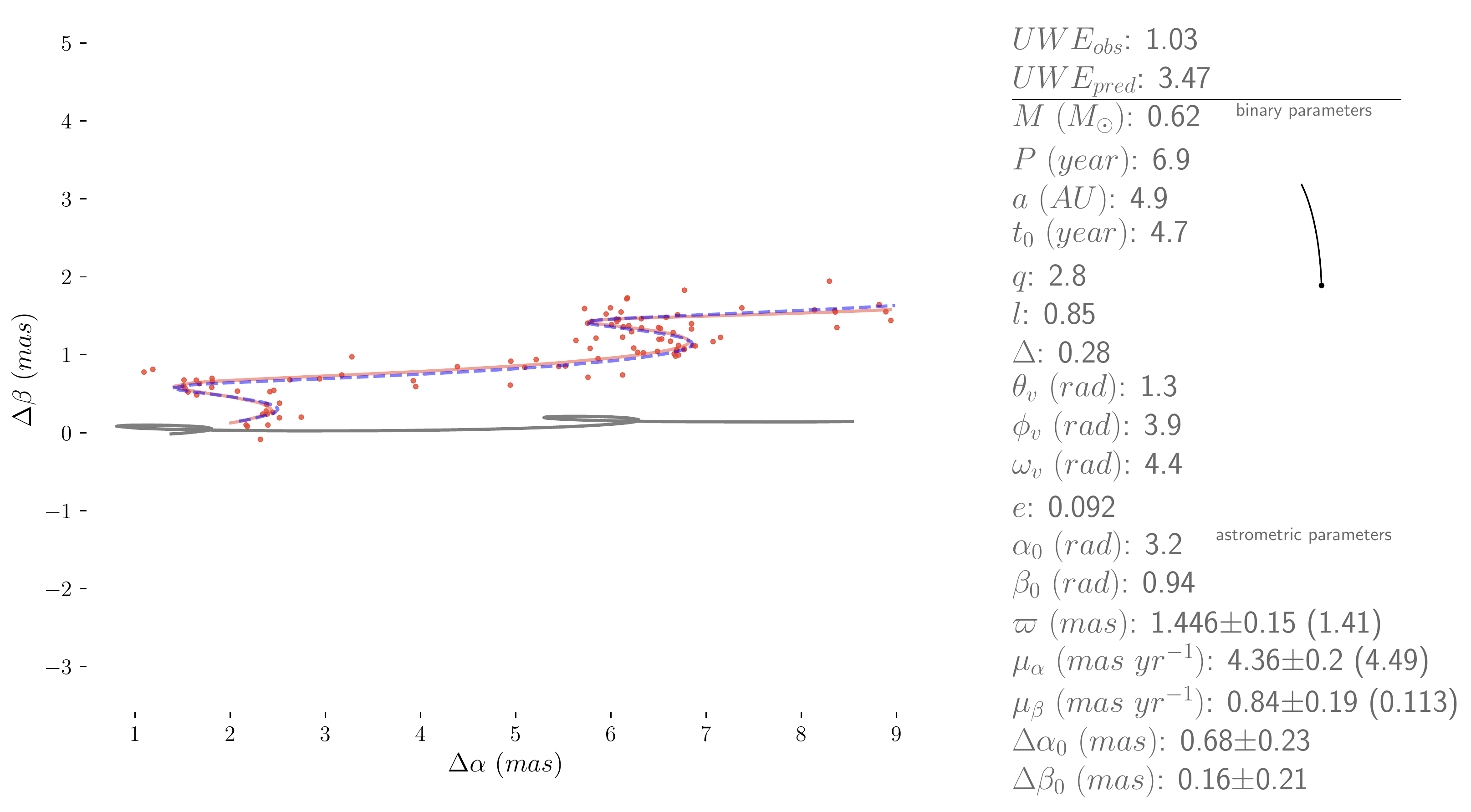}}

    \subfloat[]{\includegraphics[width=0.7\linewidth]{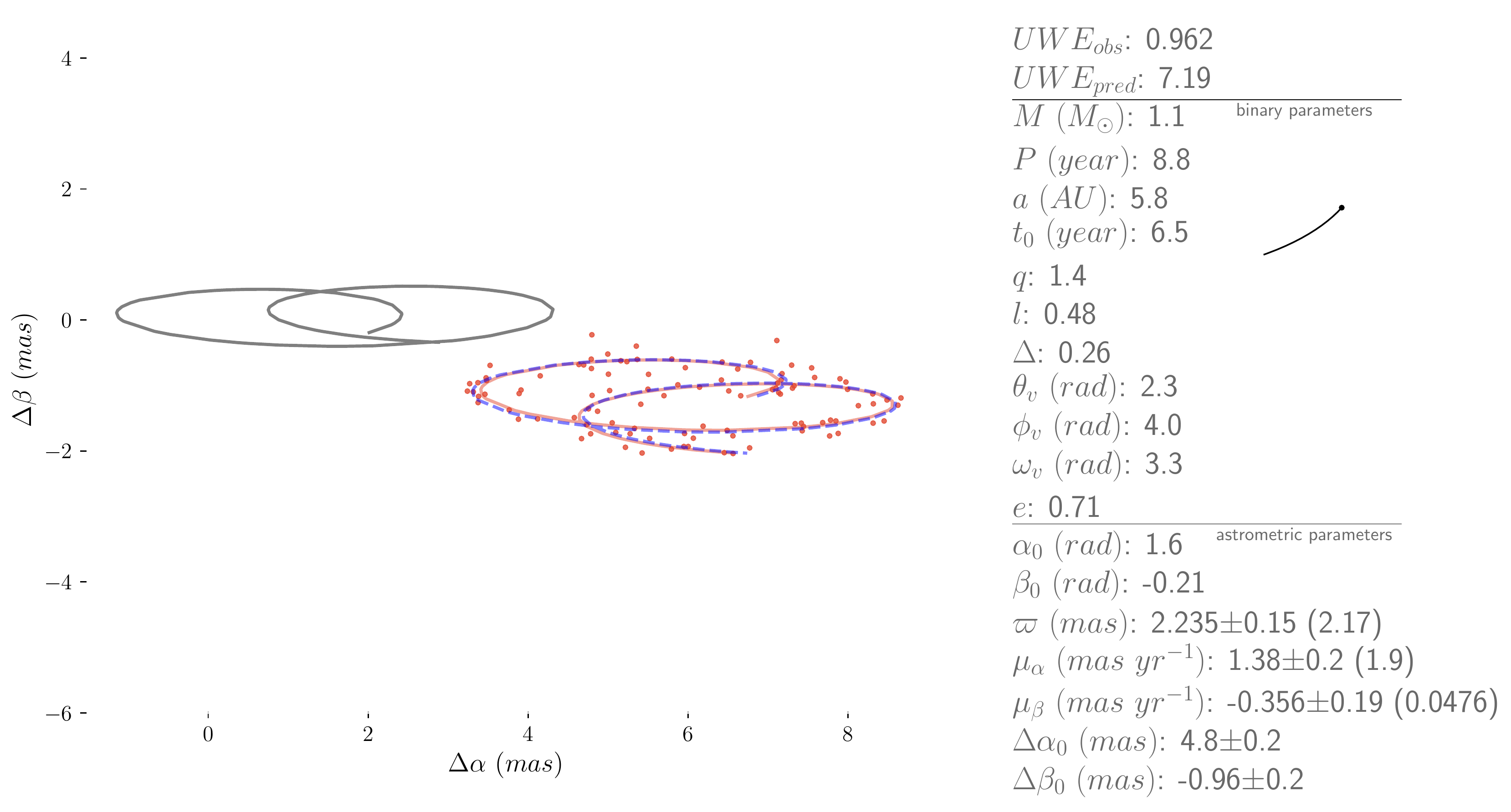}}
\caption[]{(continued)}
\end{figure*}

\begin{figure*}
\centering
\includegraphics[width=\textwidth]{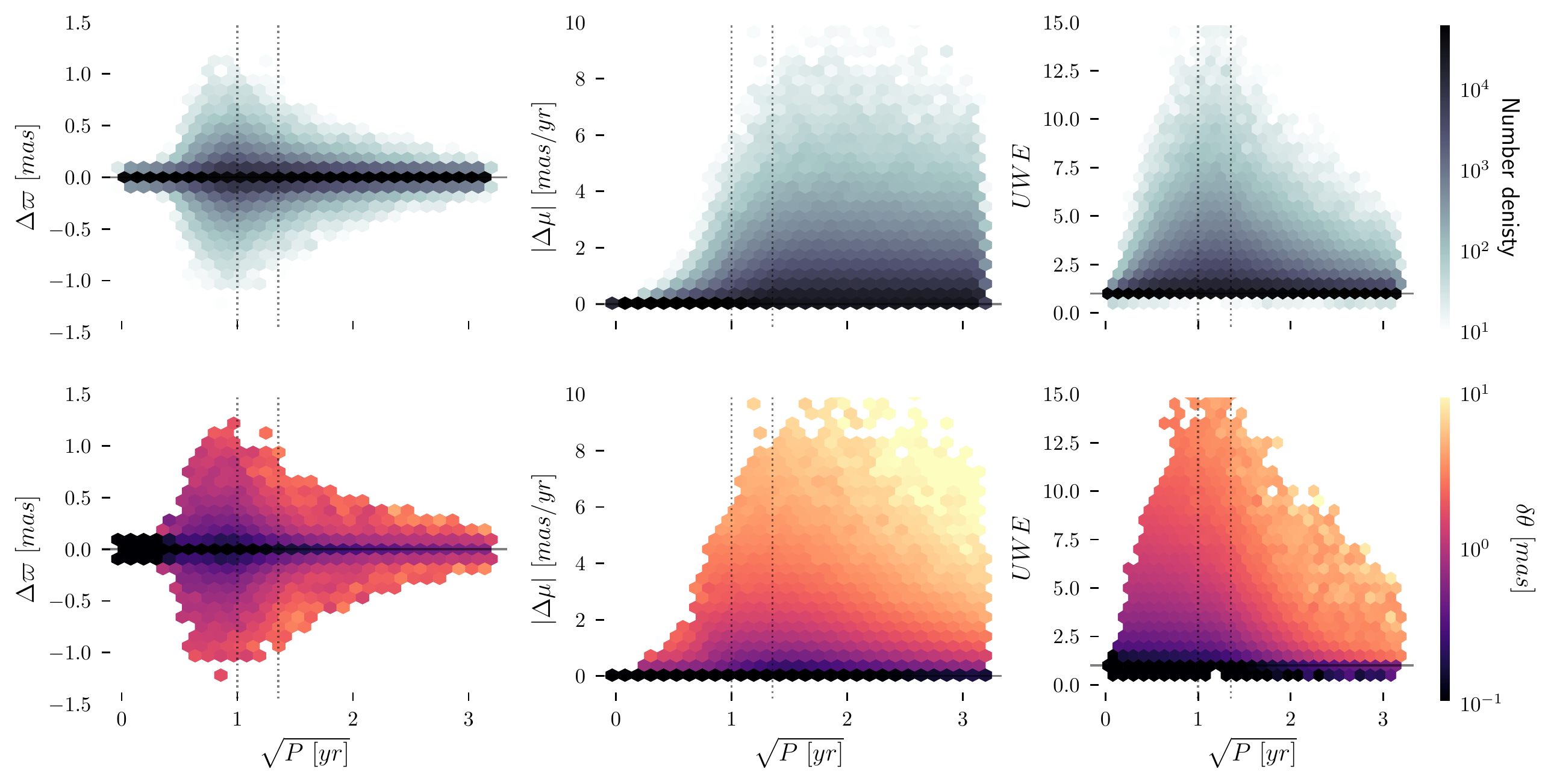}
\caption{Comparison of the parallax and proper motion deviations, and the observed UWE, as a function of period. The $x$-axis is expressed as $\sqrt{P}$ such as to have uniform density of samples (see table \ref{tabRandom}). In the upper row, figures are coloured by number density, and in the lower by the median $\delta \theta$ which is a rough representation of the magnitude of the binary contribution. The vertical dashed lines show periods of 1 year and of 22 months. We see that parallax is most affected by systems with a binary period of $\sim 1$ year, even if the effect of the binary is modest ($\delta \theta \lesssim 1$ mas). Proper motion can be affected by binaries of periods $\gtrsim 1$ year and the effect increases for more significant binary motion. Finally the observed astrometric error, as expressed through the $UWE$, also peaks towards shorter periods, though less starkly than the parallax, and scales with the binary contribution.}
\label{period}
\end{figure*}

\begin{figure*}
\centering
\includegraphics[width=\textwidth]{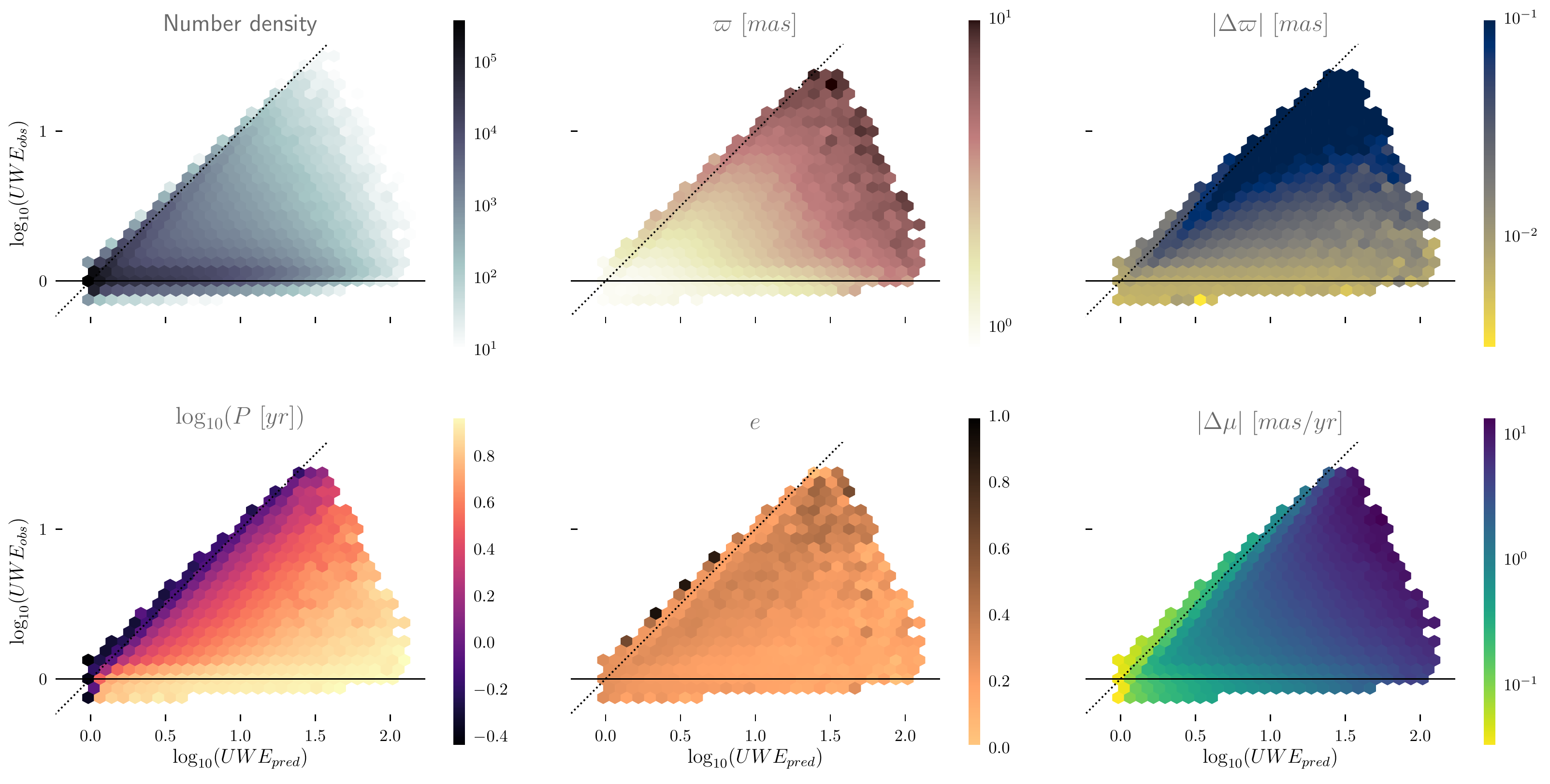}
\caption{Comparison of the predicted UWE (via section \ref{sec:analytic}) and the UWE inferred from our mock observations. The dashed diagonal line shows a $1:1$ correlation and the horizontal denotes an observed $UWE$ of 1. Colour shows the median value of the specified paramter in each bin (save for the number density plot).}
\label{uwe}
\end{figure*}

\begin{figure*}
\centering
\includegraphics[width=\textwidth]{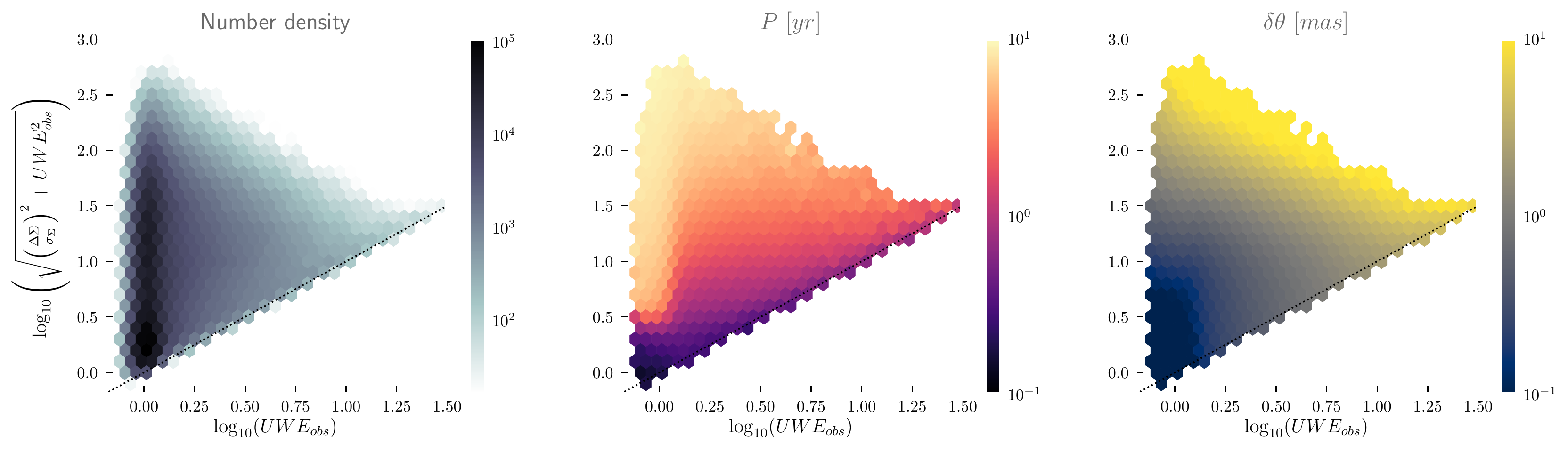}
\caption{Comparison of the total variation (the sum of squares of all parameter deviations normalised by their errors) including $UWE$ to the $UWE$ alone. A $1:1$ dotted line is shown for reference. Systems where most of the error is translated to $UWE$ lie on or near this line, whilst others where significant binary motion has resulted in large deviations of the inferred parameters (but small corresponding UWE) are well above this line. We show the number density of all measurements, and the median period and $\delta \theta$.}
\label{sponge}
\end{figure*}

We can generate mock observations by calculating the position of the centre of light at a series of times. For significant binary separations, this will deviate from the single body orbit, thus it is of interest to ask how well we might fit a single body orbit to the observed path and how far our fit may be from the true parameters.

\begin{table}
\begin{center}
\begin{tabular}{c c c}
 variable & description & distribution \\
 \hline
 $\varpi \ [mas]$ & Parallax & $10^{S[-0.05,0.17,0.36]}$ \\
 $\alpha_0 \ [rad]$ & Azimuthal position (t=0) & $2 \pi \cdot U[0,1]$ \\
 $\beta_0 \ [rad]$ & Polar position (t=0) & $\sin^{-1}(U[-1,1])$ \\
 $\mu_{\alpha} \ [mas/yr]$ & Azimuthal proper motion & $N[-1.6,7.6]$ \\
 $\mu_{\beta} \ [mas/yr]$ & Polar proper motion & $N[-3.0,7.9]$ \\
 \hline
 $l$ & Binary luminosity ratio & $U[0,1]$ \\
 $q$ & Binary mass ratio & $l\cdot10^{N[0,\frac{1}{2}]}$ \\
 $P \ [yr]$ & Binary period & $10 \cdot U[0,1]^2$ \\
 $t_0$ & Time of binary periapse & $P \cdot U[0,1]$ \\
 $M \ [M_\odot]$ & Mass of bright companion & $\frac{1}{2} (1-U[0,1])^{-0.77}$ \\
 $e$ & Binary eccentricity & $U[0,1]^2$ \\
 $\theta_v \ [rad]$ & Polar viewing angle & $\cos^{-1}(U[-1,1])$ \\
 $\phi_v \ [rad]$ & Azimuthal viewing angle & $2 \pi \cdot U[0,1]$ \\
 $\omega_v \ [rad]$ & coord. projection angle & $2 \pi \cdot U[0,1]$ \\
 \end{tabular}
\end{center}
\caption{Parameters and distributions used to generate the mock observations. The first five define the single body astrometric motion and the others define the binary motion. Here $U[a,b]$ represents a uniformly drawn random number between $a$ and $b$, $N[\mu,\sigma]$ represents a number drawn from a normal distribution with mean $\mu$ and width $\sigma$ and $S[\mu,\sigma,\varsigma]$ is a draw from a split-normal distribution with mode $\mu$, and width $\sigma$ below the mode and $\varsigma$ above.}
\label{tabRandom}
\end{table}

As an exploratory exercise, we have done this for two million systems, for which we have chosen the parameters of the binary based on the distributions listed in table \ref{tabRandom}. For simplicity, we work in coordinates aligned with the ecliptic plane $(\theta,\phi)$, though as long as we are consistent we are free to use any angular coordinate system.

The distributions of these parameters have been chosen to be both representative of real data and relatively simple. In some places, a balance has been struck between the two. For example, the distributions of $\varpi$, $\mu_\alpha$ and $\mu_\beta$ are taken from simple fits to 1 million random {\it Gaia} sources (with parallax over error greater than 15), whilst the angular positions are chosen to be uniform on-sky -- whereas in reality stars are much more clustered in the plane of the Milky Way and looking towards the galactic centre. 

The masses of the brightest star are taken from an initial mass function proportional to $M^{-2.3}$, limited to stars above $0.5 M_\odot$ (a range in which most IMFs converge). We initially experimented with an empirical period distribution of binaries from \citep{Raghavan10}, though this peaks at a period of $\sim100$ years, at which binary motion is negligible within {\it Gaia}’s temporal baseline (and also orbital separation is large enough that the sources may be independently resolvable depending on parallax). Instead we limited our period to 10 years (effectively limiting the binary separation to a few mas for our parallax distribution), and chose a distribution that favoured short periods and resembles the \citealt{Raghavan10} distribution if curtailed at 10 years, a range containing around $20\%$ of all binaries. For simplicity, we use a uniform distribution of luminosity ratios, and given that we would in general expect luminosity and mass ratio to correlate used this to inform the mass ratio -- choosing a value of $q$ log-normally distributed around $l$. Thus, $q$ will generally be close in value to $l$, but also tend to be slightly larger (luminosity normally scales strongly with mass) and have a wide spread that can encompass dark-massive companions and low-mass bright giants. The eccentricity was chosen to produce more circular orbits than highly eccentric ones. Finally, $t_0$, $\theta_v$, $\phi_v$ and $\omega_v$ are all parameters we would truly expect to be isotropic.

The sample of generated systems is intended to represent actual systems only in a loose sense -- the focus being on spanning the parameter space with a sensible distribution, not on recovering detailed statistics of actual binaries. If we were really inclined to scale up the proportion of binaries of these properties to the whole sample of observed stars in a survey such as {\it Gaia}, it would be contingent on estimating the fraction of all stars which are unresolved binaries with periods less than 10 years (correcting for other factors would likely only change the results by a factor of a few).

For each system, we calculate the position of the centre of light at 100 times, randomly spaced over a 22 month period (a rough approximation to the {\it Gaia} survey) - with an added astrometric error of $\sigma_{ast}=0.2 \ mas$, distributed isotropically on-sky. To simulate observations, we can fit single-body astrometric solutions to the sample of generated mock observations, and via linear least squares, we can find best fits and errors on $\Delta \alpha_0$, $\Delta \beta_0$, $\mu_\alpha$, $\mu_\beta$ and $\varpi$ (details of these fits are given in appendix \ref{ap:linleast}).


\subsection{Results from mock observations}

Fig.~\ref{binaries} shows a sample of eight astrometric mock observations, including their parameters. These are split into systems with a binary period less than the observing time of 22 months (left) and those with longer periods (right). All systems shown have significant binary motions, and many show substantial deviation from their centre of mass motion. However, not all of them have large $UWE$, as variation, particularly in proper motion can mimic the effect of the binary at long periods, and at short periods binary deviations can act as extra astrometric noise and just increase variance in the astrometric fit.

The top left system in Fig.~\ref{binaries} is one of a small but significant minority in which binary motion at a period close to one year enlarges (or in other cases contracts) the parallax ellipse and changes the inferred parallax significantly. Some binary motion is not easily approximated by parallactic motion - for example, in the third figure on the left, a binary period of almost exactly half a year gives a smooth well-behaved curve, but one impossible to fit well with a single body astrometric fit. At the same time, some binaries with long periods and large on-sky deviations are fit very well by the model, which translates their on-sky motion to erroneous proper motion.

\subsection{Distribution of binary deviations}

The parameter space over which we have sampled binaries is large (12 dimensional) and for real observations could feasibly be larger, including information about scanning laws and variable errors. Thus, the only conclusions we can draw from the mock data are about the large scale distributions, particularly about the magnitude of departures from the true astrometric solution for the c.o.m. motion and how this depends on binary parameters.

In Fig.~\ref{period}, we compare the period of binaries to the shift in inferred parallax, $\Delta \varpi = \varpi - \varpi_{true}$, total proper motion, $|\Delta \mu| = \sqrt{(\mu_\alpha-\mu_{\alpha,true})^2 + (\mu_\beta-\mu_{\beta,true})^2}$, and the goodness of fit as characterised by observed $UWE$ (equation \ref{uwedTheta}). We show the number density of all our mock observations (top row), and the distribution compared with reference to $\delta \theta$ (bottom row) -- a close proxy for the magnitude of the contribution of the binary.

Starting in the left hand column we see that the vast majority of systems have small and likely imperceptible parallax deviations, but some can be shifted by $1 \ mas$ or more. As we might expect, parallax shift is only significant for systems with periods close to a year, a relationship that would likely become tighter for a longer observing period. Those systems with large $|\Delta \varpi|$ tend to have a significant binary component ($\delta \theta \sim 1 \ mas$), but it is not the case that the most extreme binaries give the largest shift.

Unlike the parallax, the proper motion deviations (middle column) can be large ($\lessapprox 10 \ mas/yr$)  for any period longer than about a year, and the most extreme binaries tend to provide the largest $\Delta \mu$.

Finally the $UWE$ (right hand column) also peaks at periods close to a year, but can be significant in systems with any but the smallest period (note however that short period systems at small distances can provide significant $UWE$ but are lacking in our mock sample). For a fixed period, higher $\delta \theta$ corresponds to higher $UWE$. Few systems at 10 year periods have significant $UWE$, but there it's clear that $UWE>2$ can still occur even in systems with $P>10 \ yr$.

$UWE$ scales linearly with parallax -- and thus closer systems can have significantly larger values. The highest parallaxes used in the mock observations are around $10 \ mas$. For systems in the local vicinity of the Sun, parallaxes could reach 100's of $mas$ and thus for the same systems, the signal could be very large, or alternatively the magnitude of binary deviations could be an order of magnitude smaller and still detectable.

\subsubsection{Comparing to predicted UWE}

In Fig.~\ref{uwe}, we compare the predicted $UWE$ as calculated via the methods in Section \ref{sec:offset} to that we find from fitting to the mock data. Looking at the number density, we see that the predicted $UWE$ is effectively an upper limit on the observed $UWE$, with all systems falling on or below the 1:1 line. 

It is systems with low periods, peaking near 1 year, for which the predictions and observations agree well (lower periods may still be accurate, but both the predicted and observed signals are very close to 1). 

As might be expected, closer objects (larger parallax) have larger predicted and observed $UWE$. For a given $UWE_{pred}$, closer objects tend to have a higher $UWE_{obs}$, which is a selection effect on period (for a fixed $\delta \theta$, further systems must be wider binaries).

The eccentricity distribution is relatively flat, but there are a few interesting features to note. Given the higher number of low eccentricity orbits, we would expect these to dominate across the rest of the plot, but interestingly in the intermediate region ($1<UWE_{obs}<UWE_{pred}$ we see an overdensity of eccentric systems. The reason for this is that the information content of an eccentric orbit is syncopated, the slow motion around apoapse is about equally informative as the fast pericentre passge -- thus, for long period orbits the observed $UWE$ can still be relatively high if the short observing window overlaps with pericentre passage. This is more clearly seen in Appendix~\ref{ap:Psep}, where we separate this plot by period.

Well predicted $UWE$s tend to have a higher $\Delta \varpi$, but this is mostly due to the fact that their orbital period distribution overlaps with 1 year. 

More tellingly, the well predicted orbits tend to have low $\Delta \mu$ - showing that the effect of the binary tends to either be represented in the $UWE$ or in extra proper motion, but not both.

\subsubsection{Magnitude of deviations}

We have seen a few examples where the binary contribution can be ``absorbed" into the astrometric solution, and not show itself directly in $UWE$. In Fig.~\ref{sponge}, we explore this by comparing the total deviation from the true parameters, weighted by their errors
\begin{equation}
\frac{\Delta \Sigma}{\sigma_\Sigma}=\sqrt{\left(\frac{\Delta \varpi }{\sigma_\varpi}\right)^2 + \left(\frac{\Delta \mu_\alpha }{\sigma_{\mu_\alpha}}\right)^2 + \left(\frac{\Delta \mu_\beta }{\sigma_{\mu_\beta}}\right)^2 + \left(\frac{\Delta \alpha_0 }{\sigma_{\alpha_0}}\right)^2 + \left(\frac{\Delta \beta_0 }{\sigma_{\beta_0}}\right)^2}
\end{equation}
to the $UWE$. This quantity goes to zero along the $1:1$ line in the plot, and thus systems on this line have most of their error dominated by $UWE$, whereas significantly above the line most of the total error is absorbed within the fit.

Comparing the first and second panels, we can see that the low $UWE$ ($\lessapprox 1.2$) systems can be split into two major groups -- short period systems with smaller parameter error, and very long period binaries completely dominated by parameter error. Looking to the remainder of the middle panel, we see that intermediate period systems vary greatly in the relative contribution of errors, but interestingly contours of constant period agree well with constant total error -- i.e. the period is a good predictor of total error, but not whether it will be absorbed into the fit or the $UWE$. Finally looking at the binary contribution, as described by $\delta \theta$, we see that both $UWE$ and total error increase with more significant binaries.

This behaviour complicates the simple interpretation of a single object and what can be inferred from its $UWE$. The analytic prediction, working from an observation back to the properties of the binary, will give a lower limit on the size of the orbit/mass of the components -- but depending on unseen factors that may be a lower limit by some small percentage or orders of magnitude. For a large population, this suggests that $UWE$ will be a relatively robust measurement of binarity, though again an underestimate. It is possible that more information about the binary can be extracted by comparing to the covariance of errors in the parameter estimation ($UWE$ being effectively the collapse of these variances and covariances to a single scalar quantity).

\section{Real observations}
\label{sec:observe}

The importance of these short period binaries on astrometric observations can be split into two cases. In the first, they are a blessing, giving us a new method for identifying binary systems -- imperfectly but potentially in huge numbers, or reliably across populations. In the second, they are a nuisance, biasing a small fraction of our sample with no clean or universal way to account or adjust for them.

\subsection{Binary identification}

\begin{figure}
\includegraphics[width=\linewidth]{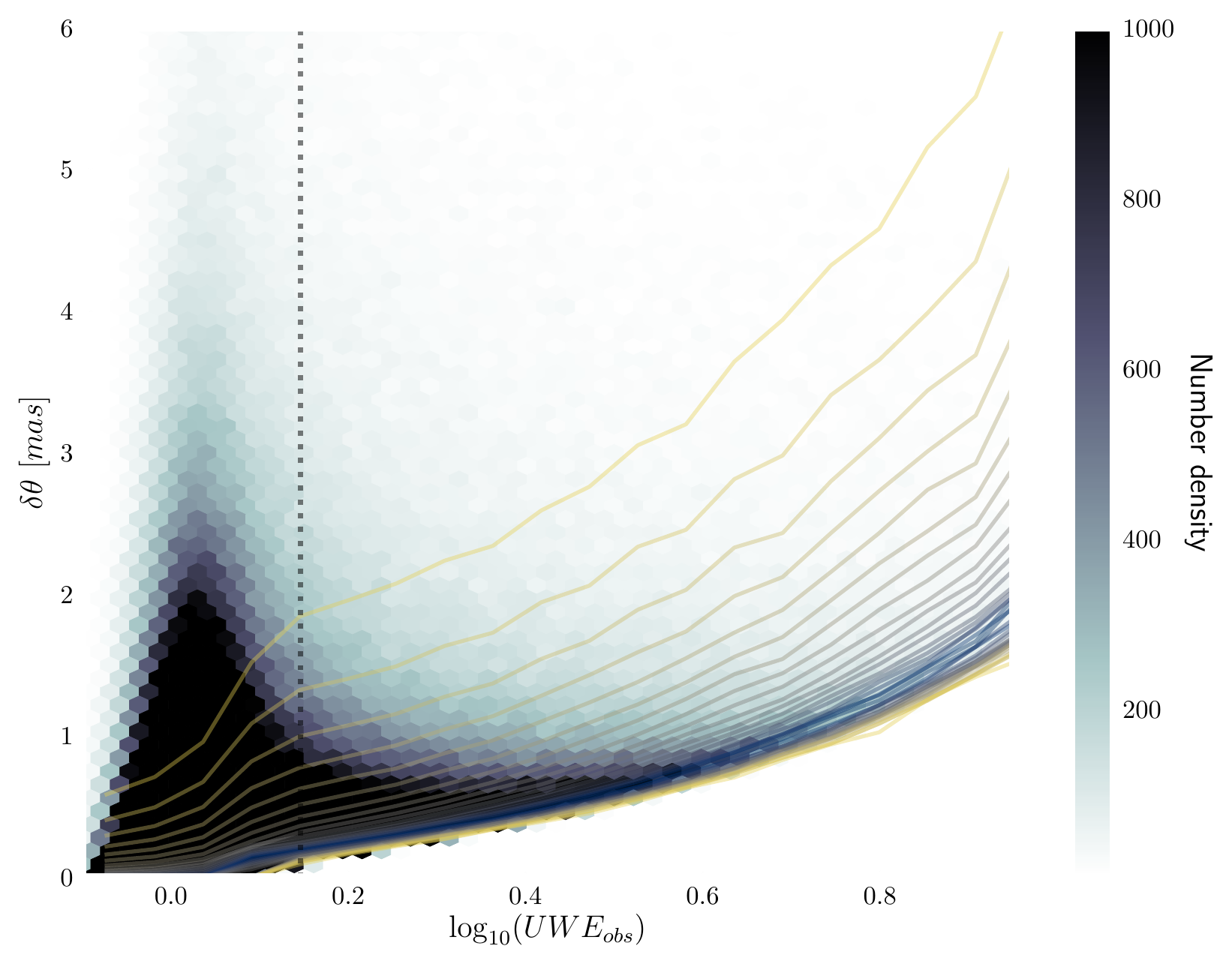}
\caption{Comparison of the observed $UWE$ to the true $\delta \theta$. Though the spread is large, above $UWE$s of around 1.4 (dotted line) there are negligible systems which do not contain a significant binary. For a given $UWE$ (observable), we might ask whether we can constrain the binary properties. Confidence intervals of $\delta \theta$ for a given $UWE$ are shown at $5\%$ intervals, ranging from $95\%$ (yellow) to $5\%$ (blue) - showing that we can predict with relatively confidence a value of $\delta \theta$ (within a factor of two) but that the tail of the distribution skews to much higher values.}
\label{observe}
\end{figure}

For any observed astrometric system, we can measure the $UWE$. The question is then whether we can reliably convert this to an inference about the presence and properties of a possible binary.

Fig.~\ref{observe} shows, as a function of the observed $UWE$ of our mock observations, the true $\delta \theta$. There is a wide spread in $\delta \theta$ for any given $UWE$, but higher $UWE$ the majority of systems do lie along a relatively tight relation (within a factor of two of the prediction of equation \ref{uwedTheta}) and, perhaps even more informatively, negligible systems lie beneath that line. Thus, according to this data observed $UWE$s of above $\sim$1.4 could reliably be inferred to correspond to binary systems, and for a population the magnitude of these binaries well estimated. For an individual system, we can only make a probable estimate of $\delta \theta$, and there is always the possibility of a wild underestimate.

If the parallax is well constrained, $\delta \theta$ can be converted to a physical separation between the c.o.l. and c.o.m. and if $\Delta$ can be estimated this can be further translated to the true binary separation. Alternatively, if the period and phase are known, we can make a much more exact estimate of $UWE$ for a given system (see Appendix~\ref{ap:anLong}) and thus comparing to observations $\Delta a$ can be precisely characterised.

This ignores any other sources of erroneous $UWE$ that may exist in the data set - either due to systematic errors, occasional oddities or other astrophysical sources. When applying this metric to real datasets the precision of measurements of $UWE$ will need to be tested explicitly to make inferences about individual objects. It may also be the case that other observed quantities, such as radial velocities or error on astrometric parameters (possibly including the covariances) can further help delineate and characterise binary systems.

\subsection{Binary contamination}

\begin{figure}
\includegraphics[width=\linewidth]{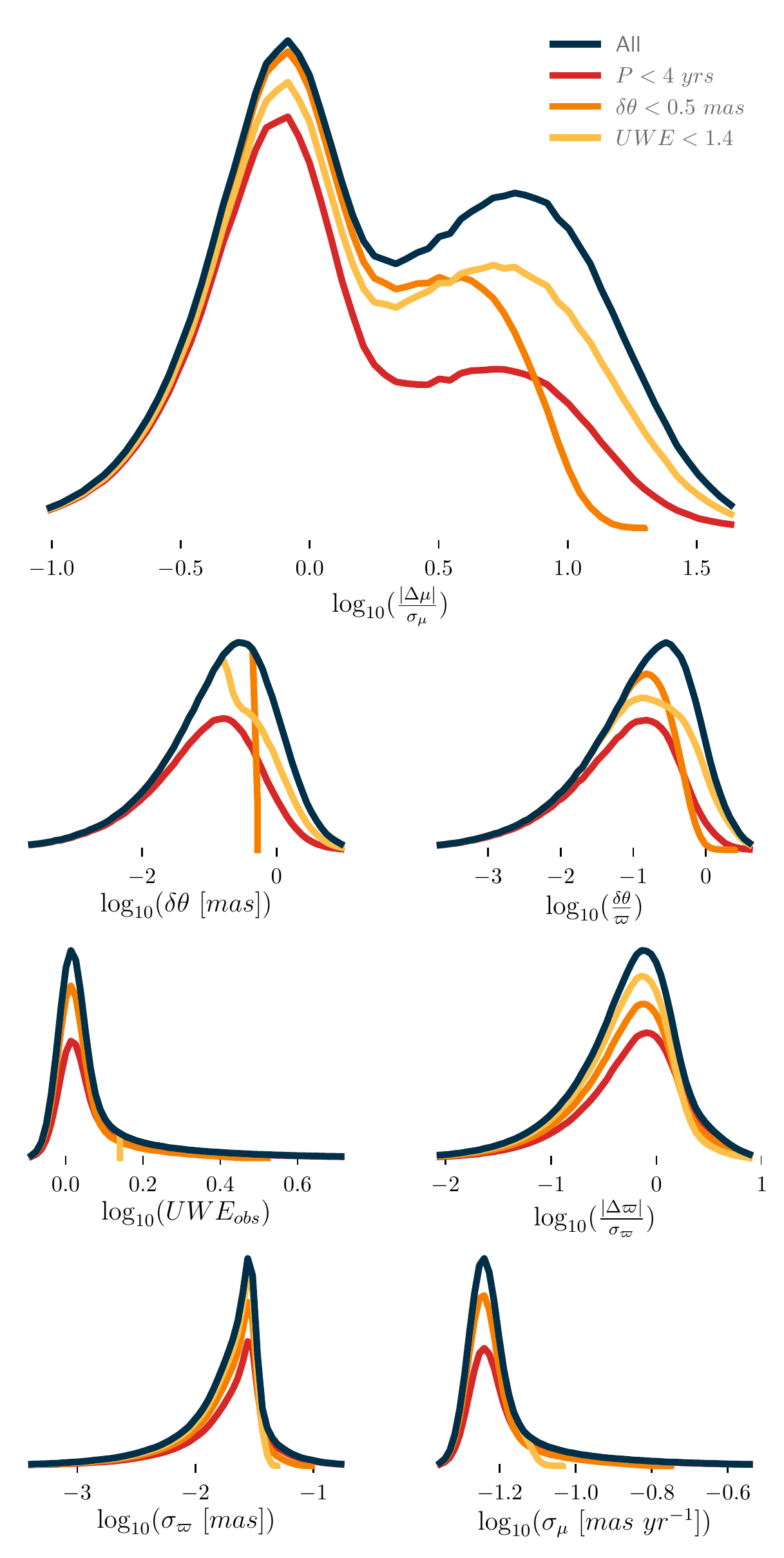}
\caption{The distribution of our mock observations, as a function of the observed deviations and errors. As well as the total sample \textit{(black)}, we show subsets with $UWE_{obs}<1.4$ \textit{(yellow)}, $\delta \theta < 0.5 \ mas$ \textit{(orange)} and $P < 4 \ yr$ \textit{(red)}. All y-scales are linear.}
\label{hists}
\end{figure}

In Fig~\ref{hists}, we show the deviations and errors observed in our mock sample of 2 million binaries. Separating by period, $\delta \theta$ and $UWE_{obs}$ illuminates which systems fill out the total distribution (black).

The largest panel shows the distribution of proper motion anomalies, with a clear bi-modal behaviour - with mostly longer period systems having significant binary-induced proper motion, whilst for most short period systems the proper motion signal is consistent with noise. A cut on $UWE$ does not differentiate these two families, whilst a cut on $\delta \theta$ puts a strong upper limit on $\frac{|\Delta \mu|}{\sigma_\mu}$.

An $UWE$ of 1.4 or below has been suggested as benchmark for removing binary contaminants. However, such a cut (yellow) still leaves around half of the binaries with $\delta \theta>0.5$ and the majority of systems which have proper motion which have been skewed by multiple $\sigma_\mu$. This sample does exclude the highest values of $\sigma_\varpi$ and $\sigma_\mu$ suggesting that they are well fit, simply erroneously so. As we have stated before, this is the impact of binaries with periods a factor of a few times the observing period, for which the partial binary orbit mimics proper motion.

Smaller binaries, with lower values of $\delta \theta$ can still have significant $UWE$ and cause large errors in parallax and proper motion. These account for almost all of the systems with a small $|\Delta \mu|$ and none of the systems with $|\Delta \mu| \gtrsim 10$. Shorter period binaries (red) can still have significant $\delta \theta$ and $UWE$, and as a population have the highest $|\Delta \varpi|$ (as we would expect given that this bracket covers the crucial 1 year binary period). Significantly fewer of the high proper motion anomaly systems have short periods. 

Before moving on from this plot, it is interesting to discuss how it would change if we had a longer observation interval. This would raise the period above which binary motion could be disguised as proper motion - narrowing the right hand peak in $\frac{|\Delta \mu|}{\sigma_\mu}$ and moving it to higher values.

Denoting the fraction of all stars which are in binaries with periods less than 10 years as $\nu$, we can make some rough estimates for the degree of binary dilution we should expect to see in astrometric samples, with $UWE<1.4$ \&
\begin{itemize}
\item $\delta \theta > 0.2 \ mas$: $30\nu \%$
\item $\delta \theta > 1 \ mas$: $8\nu \%$
\item $\Delta \varpi > \sigma_\varpi$: $20\nu \%$
\item $\Delta \varpi > 0.1 \ mas$: $0.6\nu \%$
\item $\Delta \mu > 2\sigma_\mu$: $40\nu \%$
\item $\Delta \mu > 1 \ mas \ yr^{-1}$: $4\nu \%$.
\end{itemize}
This is of course dependant on survey length. Again, we are using {\it Gaia} DR2's 22 months here -- longer baselines will lower these percentages, though they will also be able to detect deviations in systems with $P>10 yr$.

$UWE<1.4$ may still be a sensible or useful delimeter - but any such cut will let through a fraction of binaries, some of which will be significantly affected by their binarity. Depending on the case at hand, these may have little to no impact, or results may be skewed by either large numbers of small but significant binaries, or the very occasional extreme case. For example, though the shifts to $\varpi$ are generally small, this measure is necessary to calculate the absolute magnitude of the star, and thus the most extreme binary contributions may change the inferred luminosity of a star significantly.

\section{Conclusions}

This paper has studied how unresolved binary systems will alter astrometric observations. 

For shorter period binaries ($\lessapprox$ the observational baseline of the astrometric survey), the motion of the centre of light leads to increased error when fitting single body 5-parameter astrometric solutions. This excess error then provides a lower limit to the on-sky angular separation of the binary, which assuming the distance is well known can be translated to physical separation and other binary properties. It is a lower limit, as there is always the possibility that some of the binary motion is translated into a shift from the true astrometric parameters (which describe the motion of the centre of mass of the system), and thus the observed noise will be lower and the fit slightly biased. It is important to note that this ignores other sources of noise and confusion, and thus anomalously high astrometric error may be observed in single star systems -- the reliability of this metric will depend on the instrument and quite likely the particular star.  This means we can confidently make observations on a population level (when random noise will cancel out and astrometric bias will dilute our results but not mask them entirely), but inferences about individual systems will require very careful interpretation and may be impossible for many systems.

Longer period systems are more likely to bias the astrometric fit. Much of this bias is soaked up into excess proper motion (and position, but this is less physically meaningful). For systems with period close to a year, it may also cause the parallax to be under- or over-estimated. This is less likely for eccentric orbits, for whom the motion around their orbit is syncopated and thus is less easily mistaken for a parallactic ellipse.

Periods significantly longer than the observational baseline (such that negligible orbital motion is observed) will just cause a constant offset of the position (thus binaries on $10+$ year orbits will have negligible impact on {\it Gaia} DR2).

We hope this work provides a window both into how astrometric observations may be affected by binaries, but also how binaries may be identified and in some case characterised from the discrepancy between their on-sky motion and a single-body astrometric fit. We explore this directly in Belokurov et al. 2020 (submitted), in which we examine how $UWE$ varies over the whole {\it Gaia} DR2 sample - identifying populations of systems which show signs of binarity and comparing to catalogues of known binaries and exoplanet hosts.

\section*{Acknowledgements}

Special thanks to Simon Hodgkin, Emily Sandford and all the members of the Cambridge Galactic Dynamics group. SK
acknowledges the support by NSF grants AST-1813881, AST-1909584 and
Heising-Simons foundation grant 2018-1030.

\bibliographystyle{mnras}
\bibliography{bib}
\bsp

\appendix
\section{The centre of light}
\label{ap:col}

Finding the ``centre of light" of two objects is a similar, though much less well defined exercise, to finding the centre of mass. It is only really a meaningful measure when the two (or more) sources are partially or completely unresolved, and has limited physical significance, being more a function of our observations than the behaviour of the system.

For point sources, and sufficiently small extended objects, objects will appear to have some finite width set by the resolving power of our instruments (and any additional sources of noise such as atmospheric turbulence) which here we will model as a Guassian point spread function (PSF) - though a similar argument could be extended to any finite width symmetric distribution. We can model this as
\begin{equation}
b(x) \propto L e^{-\frac{(x-\mu)^2}{2\sigma^2}}
\end{equation}
where $b$ is the surface brightness at some point $x$ on-sky (which can be measured in physical units or angular distances). $L$ is the intrinsic luminosity of the source\footnote{In reality the luminosity is spread over a spectrum of wavelengths and the observed brightness depends on the response function of our telescope. As we will be comparing observations made by a single instrument, we can think of $L$ as already having taken the response function into account.}, $\mu$ is the actual position of the source and $\sigma$ is the width of the PSF. Under the assumption that that the PSF is mostly dependent on the instrument, not the source, $\sigma$ should be a constant across similar observations, and for $x$ measured in angle on-sky, $\sigma$ may be approximately constant for all observations. Though this is a one dimensional distribution the arguments can easily be extended to 2D.

We are mostly interested in pairs of unresolved sources, whose total brightness at some point along the line passing through both of their positions, can be modelled as
\begin{equation}
b_\Sigma(x) \propto L_A e^{-\frac{(x-\mu_A)^2}{2\sigma_A^2}} + L_B e^{-\frac{(x-\mu_B)^2}{2\sigma_B^2}}
\end{equation}
where $(x-\mu_A)  \ll \sigma_A$ and $(x-\mu_B)  \ll \sigma_B$ for $x$ between $\mu_A$ and $\mu_B$ (i.e. between the two sources).

Thus, using the convention from section \ref{sec:offset} where $L=L_A$ and $l=\frac{L_B}{L_A}<1$
\begin{equation}
b_\Sigma(x) = L\left(1 + l -\frac{1}{2}\left(\frac{(x-\mu_A)^2}{2\sigma_A^2} + \frac{(x-\mu_B)^2}{2\sigma_B^2} \right) \right) + O(4).
\end{equation}

When this system is observed it will appear to be a single source with a brightness $L' = L(1+l) + O(2)$ at a position $\mu'$ where $\frac{db_\Sigma}{dx}=0$. Thus
\begin{equation}
\mu'=\frac{\mu_A +\frac{l}{\epsilon^2}\mu_B}{1 +\frac{l}{\epsilon^2}} +O(2)
\end{equation}
where $\epsilon=\frac{\sigma_B}{\sigma_A}.$

Working in coordinates such that $\mu_A=0$ (centred on the brighter object) and assuming that the PSF widths are the same for both objects (which is reasonable for two objects of comparable luminosity in a close binary - though may cause significant deviations in some cases) we recover the result from section \ref{sec:offset} describing the position of the centre of light of an unresolved binary:
\begin{equation}
\mu' = \frac{l \mu_B}{1+l}+O(2).
\end{equation}

\subsection{1D scans instead of 2D images}

The above argument assumes we are free to orient the direction along which we measure the brightness of the source (and thus find the maximum) but that may not always be true - for example the {\it Gaia} survey provides much more accurate astrometric measurements parallel to the direction it scans across the sky than perpendicular, and for dimmer sources it only records 1D positions. 

If the system is scanned at an angle $\phi$ to the line connecting the two sources (where we can take $\phi$ to run from 0 to $\frac{\pi}{2}$ without loss of generality) then the measured centre of light position is modulated by a factor of $\cos\phi$. This means that for scans which only resolve perpendicular to the binary no centre of light motion is detected.

In general we can assume scan directions of the binary will be roughly isotropic and thus the observed centre of light shift will be modulated by the average of $\cos \phi$ for $0<\phi<\frac{\pi}{2}$ meaning observed displacements will be reduced by a factor of $\frac{2}{\pi}$.

In very particular cases it is possible that the scans are aligned and timed such that an effectively stationary binary (i.e. one with a long period) appears to be moving significantly on-sky and might be mistaken for other forms of motion. The frequency of such objects will be highly dependant on the form of the scanning law and a large degree of chance, but can be expected to be rare.

\section{Analytic solutions for longer period binaries}
\label{ap:anLong}

The analytic deviations derived in section \ref{sec:analytic} rely on the fact that the number of observed binary orbits is $\gtrsim$ 1, and thus the average over all time will tend to the average over a single orbit.

It is possible to perform the same analysis analytically (though requiring numerical integration) for any system providing the period and phase of the orbit at some point in time are known. For the vast majority of systems this information is exactly what we would like to derive, and thus this analysis cannot be performed. However, for known binary systems where this information is available we could in theory use this to glean yet more insight into the system.

Let us first write out the trigonometric part of equation \ref{epsVectorShort} in full
\begin{equation}
\label{epsVectorLong}
\boldsymbol{\epsilon}=\frac{\varpi \Delta a}{\Omega} \frac{1-e^2}{1+e\cos\eta} \begin{pmatrix} \cos \phi - \cos \psi_v \cos \phi_v \sin^2 \theta_v \\  \sin \phi \cos \theta_v \end{pmatrix}
\end{equation}
where
\begin{equation}
\label{omega}
\Omega(\phi_v,\theta_v)=\sqrt{1-\cos^2 \phi_v \sin^2 \theta_v}
\end{equation}
is a constant throughout the orbit.

It will be useful to convert all time dependence (currently expressed in $\phi(t)$) in terms of $\eta$ such that this becomes
\begin{equation}
\label{epsVectorEta}
\boldsymbol{\epsilon}=\frac{\varpi \Delta a}{\Omega} \begin{pmatrix} \Omega^2(\cos \eta - e) - \cos \phi_v \sin \phi_v \sin^2 \theta_v \sqrt{1-e^2} \sin \eta \\  \sqrt{1-e^2} \sin \eta \cos \theta_v \end{pmatrix}.
\end{equation}

For a significant number of observations taken at uniform (or uniformly random) intervals between some $t_1$ and $t_2$ of a known binary with period $P$ which passes through periapse at $t_0$ (which we will take to be the latest periapse passage before $t_1$) we can integrate this between $\eta_1$ and $\eta_2$ satisfying
\begin{equation}
t_1 - t_0 = \frac{P}{2\pi}(\eta_1 - e \sin\eta_1)
\end{equation}
which can be solved numerically (for $\eta_2$ we can perform the same calculation substituting $t_1$ for $t_2$).

Now we can fin the time averaged position via
\begin{equation}
\langle \boldsymbol{\epsilon} \rangle = \frac{1}{t_2-t_1}\int_{t_1}^{t_2} \boldsymbol{\epsilon} dt = \frac{1}{\eta_2 - \eta_1} \int_{\eta_1}^{\eta_2} (1-e\cos\eta) \boldsymbol{\epsilon} d\eta
\end{equation}
at this point it will be useful to define a family of integrals
\begin{equation}
I_{ab}(\eta_1,\eta_2) = \int_{\eta_1}^{\eta_2} \sin^a\eta \cos^b\eta.
\end{equation}
Letting $\Delta \eta = \eta_2 - \eta_1$ and $\Delta c_n = \cos (n \eta_2) - \cos (n \eta_1)$ (and similarly $\Delta s_n$ for sines) we can write out all the terms needed for this calculation:
\begin{equation}
\begin{aligned}
I_{00}&=\Delta \eta \\
I_{10}&=-\Delta s_1 \\
I_{01}&=\Delta c_1 \\
I_{20}&=\frac{\Delta \eta}{2}-\frac{\Delta s_2}{4} \\
I_{11}&=-\frac{\Delta c_2}{4} \\
I_{02}&=\frac{\Delta \eta}{2}+\frac{\Delta s_2}{4} \\
I_{30}&=-\frac{3\Delta c_1}{4}+\frac{\Delta c_3}{12} \\
I_{21}&=\frac{\Delta s_1}{4}-\frac{\Delta s_3}{12} \\
I_{12}&=-\frac{\Delta c_1}{4}-\frac{\Delta c_3}{12} \\
I_{03}&=\frac{3\Delta s_1}{4}+\frac{\Delta s_3}{12}
\end{aligned}
\end{equation}
Note that when $\Delta \eta$ is an integer multiple of $2\pi$ all terms except $I_{00}, \ I_{20}$ and $I_{02}$ are 0 - hence why the calculation is significantly easier if we integrate only over one full orbit. For arbitrary $\eta_1$ and $\eta_2$ these can take any value and must be precalculated (though for large $\delta \eta$ all trigonometric terms will be small, leading us back to the single orbit solution).

Performing the integral over time is thus simplified to the exercise of separating out powers of $\cos \eta$ and $\sin \eta$. This gives
\begin{equation}
\langle \boldsymbol{\epsilon} \rangle = \frac{\varpi \Delta a}{\Omega \Delta \eta} \begin{pmatrix} \Omega^2\zeta - \sin \phi_v \cos \phi_v \sin^2\theta_v \sqrt{1-e^2}(I_{10}-e I_{11}) \\  \cos \theta_v \sqrt{1-e^2} (I_{10}-e I_{11}) \end{pmatrix}
\end{equation}
and thus
\begin{equation}
\begin{aligned}
|\langle \boldsymbol{\epsilon} \rangle|^2 =& \frac{\varpi^2 \Delta^2 a^2}{\Delta \eta^2} \bigg(\Omega^2\zeta^2 - 2\sin\phi_v\cos\phi_v\sin^2\theta_v \sqrt{1-e^2}(I_{10}-eI_{11})\zeta
\\ &+ (1-\sin^2\phi_v \sin^2\theta_v)(I_{10}-eI_{11})^2\bigg)
\end{aligned}
\end{equation}
where
\begin{equation}
\zeta=(1+e^2)I_{01} - e(I_{00}+I_{02})
\end{equation}
(which we have separated out only to keep the formulas from spilling out over many lines).

Performing the same analysis we can find
\begin{equation}
\begin{aligned}
\langle|\boldsymbol{\epsilon}|^2\rangle =& \frac{\varpi^2 \Delta^2 a^2}{\Delta \eta} \bigg(I_{00} e^2
\\ &+ I_{10}2e\sqrt{1-e^2}\sin\phi_v\cos\phi_v\sin^2\theta_v 
\\ &- I_{01}e(2+e^2) \Omega^2
\\ &+ I_{20}(1-e^2)(1-\sin^2\phi_v\sin^2\theta_v) 
\\ &-I_{11}2(1+e^2)\sqrt{1-e^2}\sin\phi_v\cos\phi_v\sin^2\theta_v
\\ &+ I_{02}(1+2e^2)\Omega^2
\\ &-I_{21}e(1-e^2)(1-\sin^2\phi_v\cos^2\theta_v) 
\\ &+I_{12}2e\sqrt{1-e^2}\sin\phi_v\cos\phi_v\sin^2\theta_v
\\ &- I_{03}e\Omega^2 \bigg)
\end{aligned}
\end{equation}
and thus from equation \ref{dThetaGen} we can find $\delta \theta$ exactly.

In this regime we can also find the proper motion anomaly, by averaging $\dot{\boldsymbol{\epsilon}}$ over $\Delta \eta$:
\begin{equation}
\langle \dot{\boldsymbol{\epsilon}} \rangle = \frac{1}{t_2-t_1}\int_{t_1}^{t_2} \dot{\boldsymbol{\epsilon}} dt = \frac{\boldsymbol{\epsilon}(t_2)-\boldsymbol{\epsilon}(t_1)}{t_2-t_1}
\end{equation}
It's interesting to note that while the leading order term of $\langle \boldsymbol{\epsilon} \rangle$ decays as $\Delta \eta^{-2}$ (and $\langle |\boldsymbol{\epsilon}|^2 \rangle$ tends to a constant), the proper motion only decays as $\Delta \eta^{-1}$ on average - but will be zero for any orbit harmonic with the observing period. Thus even for large $\Delta \eta$ (many observed orbits) there may still be a significant bias on proper motion.

\section{Single body motion}
\label{ap:astrometric}

The single body motion can be captured by considering the unit vector directed towards the source from the observer. If at some initial time, $t_0$, the source is at some on-sky position (azimuthal and polar angle) ($\alpha_0$, $\beta_0$), and is moving with some proper motion $(\mu_\alpha,\mu_\beta)$, then at time $t$ the unit vector from {\it Gaia} to the source obeys
\begin{equation}
\label{astrometric}
\mathbf{\hat{r}}=\left<\mathbf{\hat{r}_0} + (t'-t_0)\left(\mu_\alpha \mathbf{\hat{p}_0} + \mu_\beta \mathbf{\hat{q}_0} + v_r \frac{\varpi}{A_u} \mathbf{\hat{r}_0}\right) - \frac{\varpi}{A_u}\mathbf{b}(t')\right>.
\end{equation}
The $< \ >$ brackets denote normalisation, $v_r$ is the radial velocity (which will disappear for all but the closest, fastest-moving stars) and $t' =  t - \frac{1}{c}(\mathbf{b}(t)-\mathbf{b}(t_0))\cdot\mathbf{\hat{r}_0}$ accounts for the slight variation in light travel time due to Earth's orbit (at most a 16 minute correction). $\mathbf{b}$ is the barycentric position of the satellite at time $t$ and $p$ is the parallax (i.e. it is this term that gives the epicycle-like motion of the source as viewed by {\it Gaia} and allows us to find the parallax) and $A_u$ is one astronomical unit. Three orthogonal unit vectors describe the line of sight direction and those of increasing azimuthal and polar angle respectively:
\begin{equation}
\label{vectors}
\mathbf{\hat{r}_0}=\begin{pmatrix}\cos\alpha_0\cos\beta_0 \\ \sin\alpha_0\cos\beta_0 \\ \sin\beta_0 \end{pmatrix}, \
\mathbf{\hat{p}_0}=\begin{pmatrix}-\sin\alpha_0 \\ \cos\alpha_0 \\ 0 \end{pmatrix}, \
\mathbf{\hat{q}_0}=\begin{pmatrix}-\cos\alpha_0\sin\beta_0 \\ -\sin\alpha_0\sin\beta_0 \\ \cos\beta_0 \end{pmatrix}.
\end{equation}
All angles and angular velocities are expressed in radians.

As $\mathbf{\hat{r}}$ gives the new approximate unit vector, we can find the azimuthal and polar angles at a given time via
\begin{equation}
\label{raDecTime}
\alpha(t) = \tan^{-1}\frac{\hat{r}_y}{\hat{r}_x} \ \ \mathrm{and} \ \ \beta(t)=\tan^{-1}\frac{\hat{r}_z}{\sqrt{\hat{r}_x^2 + \hat{r}_y^2}}.
\end{equation}
This expression ignores many (normally small) effects including evolution of the proper motions, either due to acceleration of the source or projection effects, as well as radial motion and relativistic time corrections. For our mostly qualitatively arguments it shall suffice, but a fuller description can be found in \citet{Lindegren16}.

In \ref{linearise} we linearise these equations under the assumption that motion on-sky is small to give a simpler approximate description of the motion.

\subsection{Linear model}
\label{linearise}

The one-body astrometric motion (as expressed in equation \ref{astrometric}) can be linearised in the limit of small on-sky motion. We can express the expected position of the object at time $t$ as $\alpha(t)=\alpha_i +\Delta\alpha(t)$ where $\alpha_i$ is some initial reference position which the motion remains in the vicinity of. Similarly $\beta(t)=\beta_i + \Delta\beta(t)$. Note that $\Delta\alpha_0=\Delta\alpha(t_0)$ and similarly $\Delta\beta_0$ are not necessarily 0, accounting for the small offset caused by error and binary motion. We can assume that the deviations are small, except in edge cases with coordinate singularities but these can be avoided by a change of frame.

\subsubsection{Simplifying the barycentric position}

It will be most convenient here to use coordinates aligned with the Earth's orbital plane (as it is motion in this plane that translates to the observed parallactic elliptical motion) and centred on the Sun. Thus let $\alpha_i$ be the azimuthal angle ranging covering $[0,2\pi]$ and $\beta_i$ the polar angle $[-\frac{\pi}{2},\frac{\pi}{2}]$. In these coordinates the position of Earth at time $t$ (and to a good approximation any observing instrument in Earth's orbit or at an Earth-Sun Lagrange point) is
\begin{equation}
\mathbf{b}=A_u (1-e\sin\eta)\begin{pmatrix}\cos\Phi \\ \sin\Phi \\ 0 \end{pmatrix}
\end{equation}
where $e$ is the eccentricity ($=0.0167$), $\Phi$ is the phase of Earth's orbit and $\eta$ the eccentric anomaly satisfying
\begin{equation}
\cos(\Phi)=\frac{\cos\eta - e}{1-e\cos\eta} \mathrm{,} \ \sin(\Phi)=\frac{\sqrt{1-e^2}\sin\eta}{1-e\cos\eta}
\end{equation}
and
\begin{equation}
t-t_p = \frac{T_E}{2\pi}(\eta - e\sin\eta)
\end{equation}
where $T_E$ is one year and $t_p$ is a reference time at which Earth is at periapse\footnote{For example in relevance to {\it Gaia} we might use $t_p=2456662.00 \ BJD$, shortly before the beginning of astrometric observations in {\it Gaia} DR2, $t_0=2456863.94 \ BJD$}.

In general this last expression cannot be inverted but in the limit of small eccentricity we can expand it to
\begin{equation}
\eta=\tau+e\sin\tau + O(2) \ \mathrm{where} \ \tau=\frac{2\pi(t-t_p)}{T_E}
\end{equation}
which gives
\begin{equation}
\mathbf{b}=A_u\begin{pmatrix}\cos\tau - e(1+\sin^2\tau) \\ \sin\tau + e\sin\tau\cos\tau  \\ 0 \end{pmatrix}
\end{equation}

\subsubsection{Linearized motion}

The new normalised radial unit vector obeys
\begin{equation}
\label{linSimple}
\mathbf{\hat{r}}(\alpha,\beta)=\mathbf{\hat{r}_i}+\Delta\alpha \cos\beta_i\mathbf{\hat{p}_i} +\Delta\beta \mathbf{\hat{p}_i}
\end{equation}
where $\mathbf{\hat{r}_i},\mathbf{\hat{p}_i}$ and $\mathbf{\hat{q}_i}$ are the equivalent of the vectors in equation \ref{vectors} evaluated at $(\alpha_i,\beta_i)$ and are all orthogonal.

As all deviations are small the new, non-normalised, radial vector accounting for the motion of the source is
\begin{equation}
\begin{aligned}
\mathbf{r}=&\mathbf{\hat{r}_i} +  (\Delta \alpha_0 + (t'-t_0)\mu_\alpha)\cos\beta_i \mathbf{\hat{p}_i} \\ &+  (\Delta \beta_0 + (t'-t_0)\mu_\beta) \mathbf{\hat{q}_i} + v_r \frac{\varpi}{A_u} \mathbf{\hat{r}_i} - \frac{\varpi}{A_u}\mathbf{b}(t') + O(2).
\end{aligned}
\end{equation}
All but the first term on the RHS are small and thus the magnitude of this vector is
\begin{equation}
|\mathbf{r}|=\sqrt{\mathbf{r}\cdot \mathbf{r}} = \sqrt{1 + 2 \frac{\varpi}{A_u} (v_r - \mathbf{b}(t')\cdot \mathbf{\hat{r}_i} ) + O(2)}
\end{equation}
and thus the new radial unit vector can also be expressed as
\begin{equation}
\begin{aligned}
\label{linComplex}
\mathbf{\hat{r}}=&\mathbf{\hat{r}_i} +  (\Delta \alpha_0 + (t'-t_0)\mu_\alpha)\cos\beta_i \mathbf{\hat{p}_i} \\ &+  (\Delta \beta_0 + (t'-t_0)\mu_\beta) \mathbf{\hat{q}_i} +  \frac{\varpi}{A_u} ((\mathbf{b}(t')\cdot\mathbf{\hat{r}_i})\mathbf{\hat{r}_i} -\mathbf{b}(t')) + O(2).
\end{aligned}
\end{equation}

\subsubsection{Total linearised motion}

Taking equations \ref{linSimple} and \ref{linComplex} and projecting in the $\mathbf{\hat{p}_i}$ and $\mathbf{\hat{q}_i}$ directions we can express the on-sky motion of a single body as
\begin{equation}
\begin{aligned}
\label{dalpha}
\Delta\alpha(t)=&\Delta\alpha_0+\left(t-t_0-t_b\right)\mu_\alpha \\ &- \frac{\varpi}{\cos\beta_i}\left(\cos\psi + e(\sin\psi\sin\tau-\cos\phi)\right)
\end{aligned}
\end{equation}
and
\begin{equation}
\begin{aligned}
\label{dbeta}
\Delta\beta(t)=&\Delta\beta_0+\left(t-t_0-t_b\right)\mu_\beta \\ &- \varpi\sin\beta_i\left(\sin\psi + e(\cos\psi\sin\tau+\sin\phi)\right)
\end{aligned}
\end{equation}

where
\begin{equation}
t_b=\frac{A_u\cos\beta_i}{c}\left(\cos\psi-\cos\psi_0 + e(\sin\tau\sin\psi-\sin\tau_0\cos\psi_0)\right)
\end{equation}
and $\psi(t)=\Phi(t)-\tau$.

This shows the general form of parallactic motion - a linear translation from some initial displacement (e.g. the $\Delta \alpha_0 + \Delta t \mu_\alpha$ term in equation \ref{dalpha}) and a circular motion projected on-sky due to Earth's orbit (e.g. the $\varpi \sin\beta_i \sin \psi$ term in equation \ref{dbeta}). The projection effect is stark, as polar motion due to parallax goes to zero near the ecliptic plane ($\beta_i \approx 0$) and azimuthal motion approaches a coordinate singularity at the poles (though changing to another frame of reference this behaviour disappears). This projection effect is the reason that it is much more difficult to determine parallaxes of objects on the ecliptic, only one component of the motion is visible and thus the constraining power of the observations is reduced. The small factors of $e$ and $t_b$ slightly complicate this simple picture but only at the level of a few percent, thus intuition can still be gained from this linearised form.

\section{Fitting to mock observations}
\label{ap:linleast}

To simulate observations we can fit single-body astrometric solutions to the sample of generated mock observations.

Given the linearised version of the on-sky motion (equations \ref{dalpha} and \ref{dbeta}) we can write the on-sky positions as
\begin{equation}
\begin{pmatrix}\boldsymbol{\alpha}_{obs} \\ \boldsymbol{\beta}_{obs} \end{pmatrix} = \mathbf{X} \boldsymbol{\pi} + \boldsymbol{\sigma_{\Sigma}}
\end{equation}
where $\boldsymbol{\sigma_\Sigma}$ contains the error caused by the binary and by the random systematic astrometric error, normally distributed around zero with a width $\sigma_{ast}$.

We can calculate the best fitting five-parameter astrometric model, $\boldsymbol{\hat{\pi}}$, via linear least squares:
\begin{equation}
\boldsymbol{\hat{\pi}}=\begin{pmatrix}\Delta \alpha_0 \\ \Delta \beta_0 \\ \mu_\alpha \\ \mu_\beta \\ \varpi \end{pmatrix}=\left(\mathbf{X}^T\mathbf{X}\right)^{-1}\mathbf{X}^T\begin{pmatrix}\boldsymbol{\alpha}_{obs} \\ \boldsymbol{\beta}_{obs} \end{pmatrix}
\end{equation}
where $\boldsymbol{\alpha}_{obs}$ and $\boldsymbol{\beta}_{obs}$ are the vector of $N_{obs}$ (=100) mock azimuthal and polar coordinates and
\begin{equation}
\mathbf{X}(\mathbf{t},\theta,\phi)=\begin{pmatrix}
\mathbf{1}, \ \mathbf{0}, \ \mathbf{t}-\mathbf{t_b}(\mathbf{t},\theta,\phi), \ \mathbf{0}, \ \mathbf{p}_\alpha(\mathbf{t},\theta,\phi) \\
\mathbf{0}, \ \mathbf{1}, \ \mathbf{0}, \ \mathbf{t}-\mathbf{t_b}(\mathbf{t},\theta,\phi), \ \mathbf{p}_\beta(\mathbf{t},\theta,\phi)
\end{pmatrix}
\end{equation}
where
\begin{equation}
\mathbf{p}_\alpha(\mathbf{t},\theta,\phi) = -\frac{1}{\cos\theta}\left(\cos\boldsymbol{\psi} + e\left(\sin\boldsymbol{\tau}\sin\boldsymbol{\psi} - \cos\phi \right) \right)
\end{equation}
and
\begin{equation}
\mathbf{p}_\beta(\mathbf{t},\theta,\phi) = -\sin\theta\left(\sin\boldsymbol{\psi} + e\left(\sin\boldsymbol{\tau}\cos\boldsymbol{\psi} + \sin\phi \right) \right)
\end{equation}
($\mathbf{0}$ and $\mathbf{1}$ are vectors of $N_{obs}$ zeros and ones respectively, $\mathbf{t}$ are the $N_{obs}$ observing times and $\boldsymbol{\psi}$ and $\boldsymbol{\tau}$ the corresponding $N_{obs}$ values of $\psi(t,\phi)$ and $\tau(t)$).

We can calculate the observed Unit Weight Error as
\begin{equation}
\label{eq:uweObs}
UWE_{obs}=\frac{\left|\left|\begin{pmatrix}\boldsymbol{\alpha}_{obs} \\  \boldsymbol{\beta}_{obs} \end{pmatrix} - \mathbf{X}\boldsymbol{\hat{\pi}}\right|\right|}{\sigma_{ast}\sqrt{N_{obs}-5}}
\end{equation}

The corresponding errors in the parameters follow the $5$ by $5$ matrix
\begin{equation}
\boldsymbol{\hat{\sigma}}_\pi^2 = \sigma_{ast}^2 \cdot UWE_{obs}^2 \left(\mathbf{X}^T\mathbf{X}\right)^{-1}
\end{equation}
where the on-diagonal terms gives us the variance on a single parameter and the off-diagonal terms the covariances. We will express approximate errors in the parameters as the square-root of the on-diagonal terms.

\section{Predicted versus observed UWE by period}\label{ap:Psep}

\begin{figure*}
\centering
    \subfloat[]{\includegraphics[width=\linewidth]{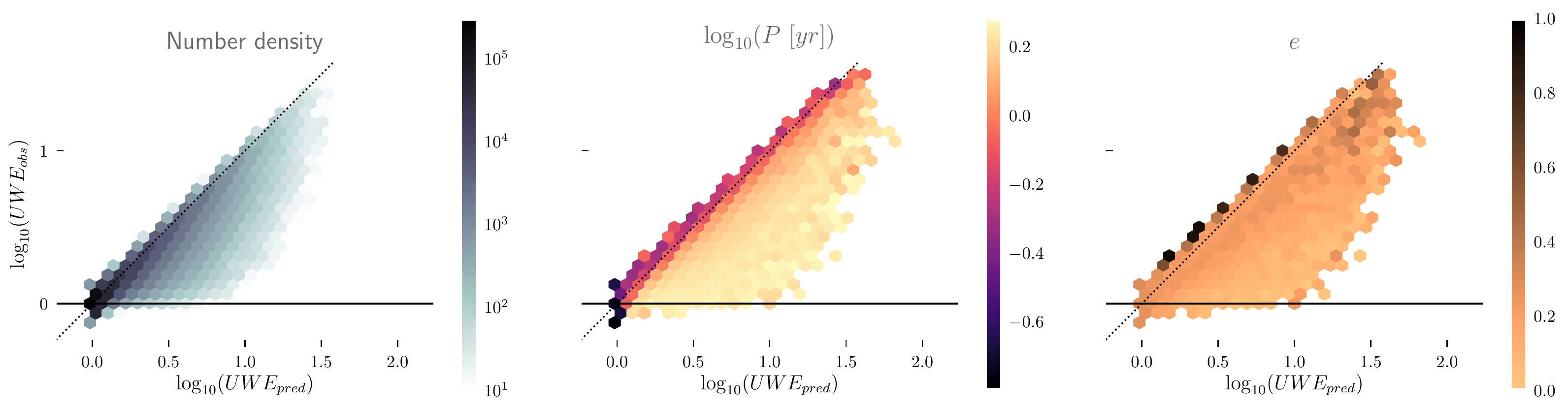}}

    \subfloat[]{\includegraphics[width=\linewidth]{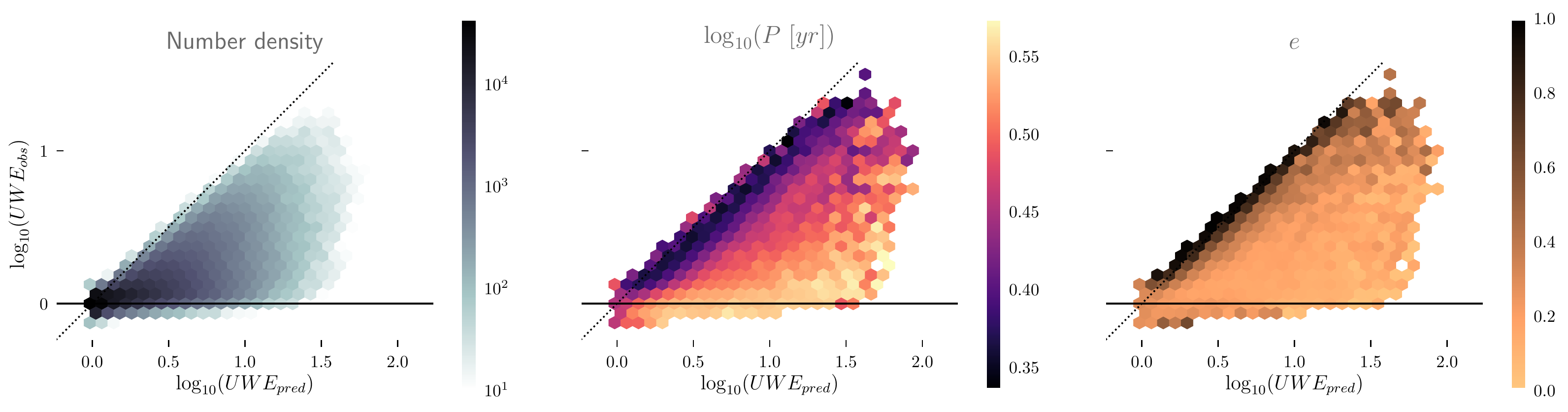}}

    \subfloat[]{\includegraphics[width=\linewidth]{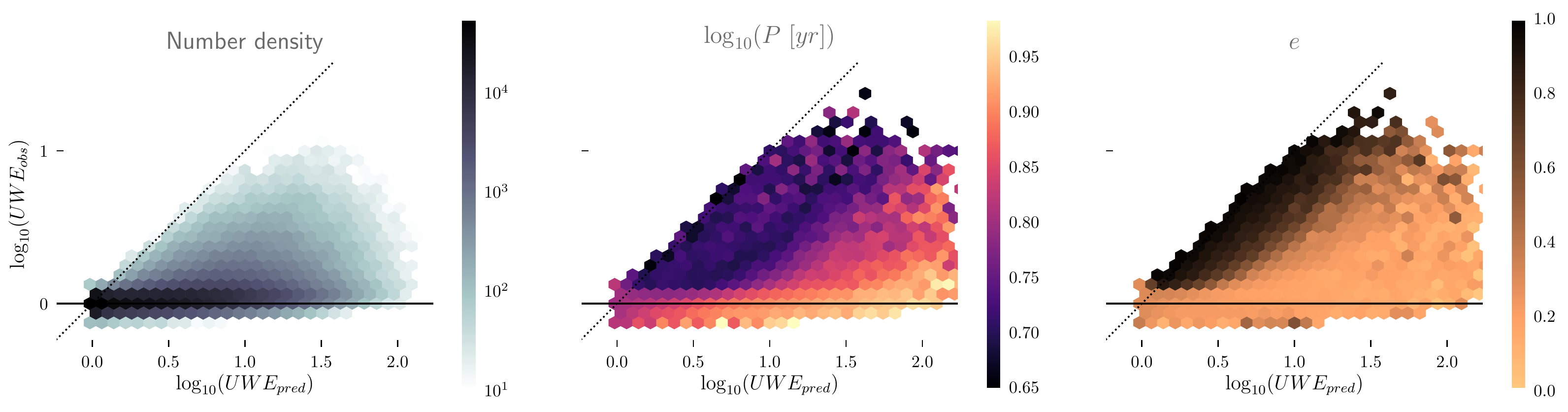}}

\caption{Predicted vs observed $UWE$ (as in figure \ref{uwe}) from our mock observations, separated by period of binary orbit. Top: $P< 2 \ yr$, Middle: $2 \ yr < P < 5 \ yr$ and Bottom: $P > 5 \ yr$. Note the changing scale of the colour bar for the median periods.}
\label{uwePeriod}
\end{figure*}

Figure \ref{uwePeriod} shows the predicted $UWE$ compared to the observed value for binaries divided into three period intervals (separated at 2 and 5 years). Now we can see very clearly the high good agreement between predictions and the mock observations for short period systems. Even for periods a few times linger than the observing baseline (22 months) the observed $UWE$ can be large.

Here we can see clearly that highly eccentric orbits, even on long periods, can have large observed $UWE$ - as though only part of the orbit is resolved if that fraction overlaps with the fast motion through periapse passage we still capture much of the total orbital motion.

\label{lastpage}

\end{document}